\documentclass[twocolumn,showpacs,preprintnumbers,amsmath,amssymb,prb,superscriptaddress]{revtex4}
\usepackage{graphicx}
\usepackage{bm}
\begin{document}

\title{Frustrated phase separation in two-dimensional charged systems}
\author{C.  Ortix}
\affiliation{Dipartimento di Fisica, Universit\`{a} di Lecce and INFN Sezione di Lecce, Via per
  Arnesano, 73100 Lecce, Italy. }
\author{J.  Lorenzana} 
\affiliation{Dipartimento di Fisica, Universit\`a di Roma 
``La Sapienza'', P.  Aldo Moro 2, 00185 Roma, Italy. }
\affiliation{SMC-Istituto Nazionale di Fisica della Materia}
\affiliation{ISC-Consiglio Nazionale delle Ricerche} \author{C.  Di
Castro} 
\affiliation{Dipartimento di Fisica, Universit\`a di Roma 
``La Sapienza'', P.  Aldo Moro 2, 00185 Roma, Italy. }
\affiliation{SMC-Istituto Nazionale di Fisica della
Materia}
 \date{\today}

\begin{abstract}
We study phase-separation frustrated (FPS) by the long-range 
Coulomb interaction
in two-dimensional electronic systems with emphasis to the
case of a metallic and an insulating phase. In the mixed
phase the system self-organizes in terms of mesoscopic inhomogeneities
of one phase hosted by the other phase. We analyze the cases
of circular drops and of alternating stripes.
As a first approximation we consider the density inside each
inhomogeneity as constant and in some cases we test 
the accuracy of this assumption by a more involved local density 
approximation.
We find that the transition from the uniform phase to the
frustrated phase-separated  phase changes order depending 
upon its geometric arrangement. 
 Contrary to what was found in three dimensional systems,
 there is no upper bound for the size of inhomogeneities. 
This difference stands on the different role of the long-range Coulomb
 interaction and screening in two and three dimensional systems. We conclude  
 that  two-dimensional systems are more prone to mesoscopic FPS.
\end{abstract}
\maketitle
\section{Introduction}
A variety of strongly correlated charged systems shows a strong tendency to
phase separation (PS).
\cite{mul92,sig93,mor99,nag83,nag98,hen98,low94,cas95b,ila00,ila01,pan01,mce03,lan02,bec02}
 Often these system are two-dimensional (2D) like the two-dimensional 
electron gas (2DEG)\cite{ila00,ila01} or quasi two-dimensional 
like cuprates\cite{mul92,sig93} or some
 manganites.\cite{mor99,sal01,per97,lar01,zha02,lar05}

Advances in local probe techniques have
revealed the mesoscopic nature of the phase coexistence. For example
cuprates  phase separate into superconducting like regions
and poorly metallic regions on the scale of $\sim 5nm ~\sim 10$ lattice
constants.\cite{pan01,mce03,lan02} The regions have round like shapes
indicating that surface energy has an important
role. 

Mesoscopic phase coexistence has also been reported in
manganites.\cite{sal01,per97,hen98,fat99,lar01,ren02,bec02,lar05}
Here the situation is quite complex. Some systems have 
large insulating/metallic clusters with fractal like
interfaces.\cite{fat99} In this case phase separation is dominated by
disorder effects stabilizing one or the other of the two phases very close
in energy. On the other hand scanning tunneling spectroscopy in 
thin films have revealed filamentary and drop like metallic/insulating
regions in the scale of tenths to thousand of nanometers 
 with smooth surfaces indicating strong surface energies. 
In both cases percolation of the metallic regions is closely
correlated to abrupt changes in transport.\cite{bec02,zha02}
Finally neutron and X-ray scattering has revealed much smaller clusters, on the
nanometer scale, in the bulk.\cite{per97,hen98,lar01,lar05}

Using a local probe in the 2DEG, Ilani 
and collaborators\cite{ila00,ila01} have shown that close to a
puzzling metal-insulator transition\cite{kra94,kra95,sim98}
 the system becomes inhomogeneous at a mesoscopic scale. In addition the compressibility close to the transition,  
departs sharply from the predictions of an homogeneous electron gas.\cite{eis94,eis92,dul03,dul00,ila00,ila01}

Mesoscopic inhomogeneities are generally expected in systems 
with a density driven first order phase-transition in the presence of
long-range forces. This phenomenon is well known in a variety of
systems\cite{seu95} ranging from neutron stars\cite{lor93prl} to
spinodal decomposition hampered by elastic forces\cite{fra99} and it is also 
related to the problem of domain formation in ferromagnetic 
systems.\cite{kit46,lan84}

In charged systems the phase coexistence phenomenon
is hampered by the long-range Coulomb interaction (LRC).  
Indeed a macroscopic charge imbalance would imply 
an electrostatic energy cost which grows faster than the volume 
in the thermodynamic limit. Thus 
it has been proposed that the system should break in domains
in order to guarantee large-scale neutrality.
\cite{lor93prl,nag83,nag98,low94,cas95b,lor01I,lor01II,lor02,mur02}
In this way the charge is
segregated over some characteristic distance but the average density
at large distances is constant. 

When the scale of the inhomogeneities
is mesoscopic, one can perform a general analysis of FPS independently of the specific short-range interaction favoring PS, in the 
same spirit of the Maxwell
construction (MC).\cite{lor93prl,nag83,nag98,lor01I,lor01II,lor02,mur02} 
 The specific short-range interaction, however, will account for different
physical situations with different coexisting phases. Due to this
mesoscopic hypothesis the inhomogeneities can be treated as 
charged classical objects. Their size and their relative distance are
determined by the competition between the LRC interaction and the
interface boundary energy. These effects determine the total ``mixing
energy'' i.e. the excess of energy with respect to the unfrustrated PS. 
To characterize the
degree of frustration  one can define a dimensionless parameter  
$\lambda$ given by the ratio of the characteristic 
mixing energy  and the characteristic
 energy gain due to PS. 

In 3D systems it
 was shown that the LRC interaction favors 
uniform phases.\cite{lor01I,lor01II,lor02} Indeed the coexistence
region shrinks as
$\lambda$ increases and the uniform phases are stabilized at
densities where the ordinary MC analysis 
would predict a PS state. In addition
 the size of the
inhomogeneities in 3D systems has been shown to satisfy a ``maximum size theorem'' 
that says that inhomogeneities can not have all linear dimensions  
much larger than the screening length.\cite{lor01I,lor01II,lor02}

 In this work we  consider mesoscopic frustrated
phase-separation for 2D electronic 
systems, that is electrons confined to a plane but subject to the 
3D Coulomb interaction in the presence of a rigid background. 
Muratov, instead, has considered the case of a $d$-dimensional
system immersed in the $d$-dimensional Coulomb interaction.\cite{mur02}
As a first approximation we neglect disorder effects and concentrate on the 
thermodynamic behavior in the clean limit.  
  We analyze in  detail the coexistence between a compressible phase 
(named ``B'') and an incompressible phase (named ``A'') corresponding
to the physically relevant case of phase separation between a metal and
an insulator. The results are easily generalizable 
to two compressible phases.\cite{note} A non-rigid background produces
peculiar effects close to an energy level crossing of the electronic 
phases\cite{lor01I,spi03,spi04} and
will be discussed elsewhere in the present context.\cite{ort06}

 We consider two different types of 
inhomogeneities: drops of one phase into the other phase and 
 alternating stripes of each phase.
We assume a uniform density inside each inhomogeneity.  This uniform density 
approximation (UDA) is relaxed in Sec.~\ref{seclda} and tested in
Appendix \ref{seckbinf} against a
more involved local density approximation (LDA). UDA is shown to be 
quite accurate for the evolution of global thermodynamic quantities. 

As in 3D systems, the LRC
 interaction stabilizes the homogeneous metallic phase at
densities where non frustrated systems would show PS .

In contrast to the 3D case, in 2D the density driven 
transition to the inhomogeneous state 
changes order depending upon the different geometric arrangement. Within
our approximation, one 
has a second-order transition to a droplet state while
the transition to the stripes geometry is always  
first-order like.

Moreover in 2D systems the size of the 
inhomogeneities is not limited by the screening length. 
As a result we find that 2D systems are more prone to mesoscopic 
FPS than the 3D systems.

\section{General Analysis}
\label{secgeneral}
As in the 3D case we assume a separation of length scales. In many models on a lattice it is found that short-range forces tend to phase separate the electrons between electron rich 
and electron poor regions.\cite{eme90c,kiv90,mor99,can91,gri91,don95,oka00} 
We assume that the
short range forces lead to a double-well form for the energy of an hypothetical  uniform phase as a function of density.
We call $f_{A}$ and $f_{B}$ the energy densities close to each minimum,
which define the bulk energy of the phases.  
 Long-range forces 
prevent large scale phase separation leading instead to domain formation.
As a simplified assumption we assume uniform densities
($n_{B}$,$n_{A}$) for the inhomogeneities of each phase (UDA) 
with sharp interfaces (soft interfaces have been
considered by Muratov\cite{mur02}). 
This charge distribution is compensated
by a rigid background of
density $n$. The UDA  will be relaxed in section \ref{seclda}. 

The free energy per unit ``volume'' of the mixed-phase reads:
\begin{equation}
\label{eq:f}
f=\left(1-\nu\,\right)\,f_{A}\left(n_{A}\right)\,+\,\nu\,f_{B}\left(n_{B}\right)\,+\,e_{el}\,+\,e_{\sigma}
\end{equation}
where $ \nu $ denotes the volume fraction of the B-phase ($V_{B}/V$),
 $ e_{el} $ represents the LRC interaction energy density and
$ e_{\sigma} $ is the ``surface'' energy density. (Here and below, in order to 
keep a common nomenclature with the 3D case,  we call
``volume'' a quantity with units of length squared and ``surface'' a
quantity with units of length). 
Due to charge neutrality the global charge density of electrons
has to compensate the charge density of the background $e n$. 
This leads to the following neutrality constraint: 
\begin{equation*}
n = \left(1- \nu\right) n_{A} + \nu\, n_{B}
\end{equation*}

To proceed further we assume specific geometries for the inhomogeneities. We
will consider the competition among the two following geometries:  
 {\it i}) drops (disks) of one phase surrounded by the other phase 
and {\it ii}) a periodic
structure of alternating stripes of the two coexisting phases.

For the drops we
divide the system in cells of radius $ R_{c}
$ enclosing one domain of the B-phase with
radius $ R_{d} $. Similarly for the stripes the cell has width $2
  R_{c}$ and contains a B-phase stripe of width $2 R_{d}$. The 
volume fraction is related to these
characteristic lengths by:
\begin{equation*}
\nu\,=\,\left(\dfrac{R_{d}}{R_{c}}\right)^{\alpha}
\,\,\,\,\,\,\,\,\,\alpha=1\,,\,2
\end{equation*} 
where for 2D systems  $\alpha=1, 2$ for the stripe geometry and the drop
geometry respectively. In the latter case the cells are slightly
overlapping with: $\,\pi R_{c}^{2}\,N_{d}\,=\,V$ ($N_{d}$ indicates the
number of cells in the system while $ V $ is the total
volume). For the stripe geometry $V=2R_{c}\,L\,N_{d}$ where $L$
indicates the length of the stripes. The total surface energy per unit
volume can be parameterized by a
quantity $ \sigma $ with dimensions of energy per unit surface
(actually length in 2D). It reads:  
\begin{equation*}
e_{\sigma}\,=\,\sigma\,\frac{N_{d}\,\Sigma_{d}}{V}
\end{equation*}  
where $ \Sigma_{d} $ is the surface of the domain interface inside one cell with $\Sigma_{d}=2\,\pi\,R_{d}$ and $\Sigma_{d}=2L$ for the
drops and the stripes respectively. For inhomogeneities of the B-phase hosted by the
A-phase, $e_{\sigma}$ can be written as:
\begin{equation}
e_{\sigma}\,=\,\frac{\sigma}{R_{c}}\,\alpha\,\nu^{\frac{\alpha-1}{\alpha}}
\label{dsurface}
\end{equation}

The operation: $A \leftrightarrow B$, $\nu\leftrightarrow 1-\nu$ was named ``phase exchange'' in
Ref.~\onlinecite{lor01I}. Within the UDA the energy should not change 
under this operation.\cite{lor01I} For the stripe geometry case the surface energy does not
depend on $\nu$ and Eq.~(\ref{dsurface}) already preserves 
the phase-exchange symmetry. On the contrary for the drop geometry  Eq.~\eqref{dsurface} is appropriate only for
small $\nu$. In fact for intermediate volume fractions the drops should
deform and for $\nu\simeq 1$ 
the B-phase should represent the host in which drops of the A-phase are
immersed. Eq.~\eqref{dsurface}, instead, 
violates this phase exchange symmetry 
because it involves drops of B-phase in both cases. We can define a ''symmetrized'' interpolating form that is correct at the  
two extremes ($\nu\simeq 0$ and $\nu\simeq 1$): 
\begin{equation}
e_{\sigma}\,=\,\frac{\sigma}{R_{c}}\,\alpha\,\left[\nu\left(1-\nu\right)\right]^{\frac{\alpha-1}{\alpha}}
\label{dsurfapprox}
\end{equation} 
 This is enough for our proposal because 
we can anticipate that drops are stable only in a narrow 
region close to $\nu\simeq 0$ and $\nu\simeq 1$ 
({\it c.f.} Fig.~\ref{ucomparison}).

The long-range Coulomb interaction energy is computed in the
Appendix~\ref{secappe} by dividing the systems in neutral Wigner-Seitz
cells. For stripes the electrostatic energy is computed 
numerically. 

The electrostatic energy contribution due to the interaction between different cells can be neglected for volume fractions close to $\nu\sim0$ and $\nu\sim1$.
In these limits, in fact, the inhomogeneities are far from each
other and the interaction becomes irrelevant.  We show below that this
is indeed a reasonable approximation in the full range of volume fractions. 
As discussed in Appendix~\ref{secappe},
we expect this approximation to be even more accurate 
for the drop geometry. Therefore for further analytical computations
we use the  approximate expression derived in Appendix~\ref{secappe} 
for both geometries:
\begin{equation}
e_{el}\,=\,\dfrac{e^{2}}{\varepsilon_{0}}\left(n_{B}-n_{A}\right)^{2}\,R_{c}\,\,\frac{8}{3}\left[\nu\left(1-\nu\right)\right]^{\frac{3}{2}}
\,\,\,\,\,\,\,\,\, \text{Drops}
\label{electrodrops}
\end{equation}
\begin{eqnarray}
e_{el}&=&\dfrac{e^{2}}{\varepsilon_{0}}\left(n_{B}-n_{A}\right)^{2}\,R_{c}\,\,2\left[\nu\left(1-\nu\right)\right]^{2}\left[-\log
\nu \left(1-\nu\right)\right]\nonumber\\ && \,\,\,\,\,\,\,\,\,\,\,\,\,\,\,\
\,\,\,\,\,\,\,\,\,\,\,\,\,\,\,\,\,\,\,\,\,\,\,\,\,\,\,\,\,\,\,\,\,\,\,\,\,\,\,\,\,\,\,\,\,\
\,\,\,\,\,\,\,\,\,\,\,\,\,\,\,\,\,\,\,\,\,\,\,\,\,\,\,\,\,\,\,\,\,
\text{Stripes} 
\label{electrostripes}
\end{eqnarray}
Here $\varepsilon_{0}$ is the static dielectric constant.
For the drop
geometry the electrostatic energy has been symmetrized similarly to the surface energy.

\begin{figure}[tbp]
\includegraphics[width=8cm]{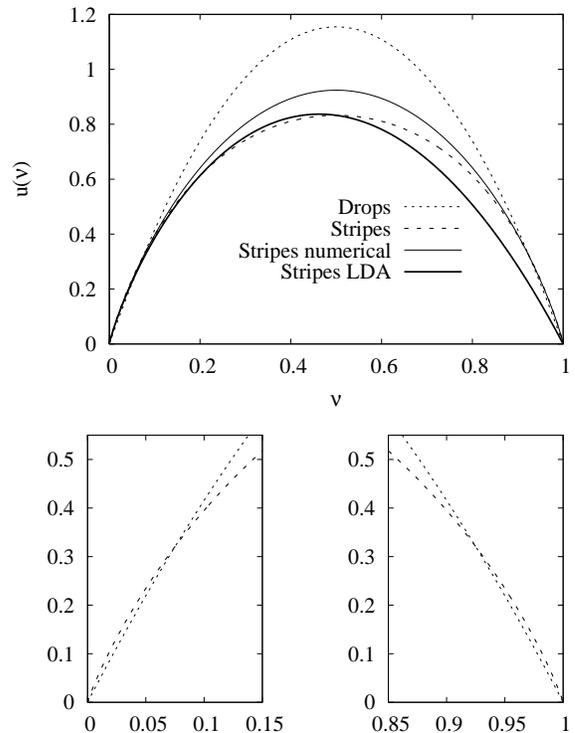}
\caption{Top: Approximate $u$ functions
  parameterizing the mixing energy for the drops [Eq.~\eqref{eq:udrop}]
  and the stripes [Eq.~\eqref{eq:ustripe}]. For the latter geometry we also provide the
  numerically evaluated expressions taking into account the
  electrostatic interaction energy
  ({\it see} Appendix\ref{secappe}), and the LDA ({\it
  see} Appendix.~\ref{seckbinf}) in order to test our
  approximations. Bottom: Expanded scale for $\nu \simeq 0$ and $\nu \simeq 1$ : the drop geometry introduces a lower mixing energy
  contribution.}
\label{ucomparison}
\end{figure}

The bulk free energies $f_{A}$ and $f_{B}$ appearing in Eq.~\eqref{eq:f} do not depend
upon $R_{c}$. The cell radius is therefore
determined by minimizing the mixing energy $ e_{m}\,=e_{el}+e_{\sigma}
$ at fixed $\nu$. For the two geometries we have:
\begin{equation}
R_{c}\,=\,\left(\frac{\sigma\,\varepsilon_{0}}{e^{2}\,
\left(n_{B}-n_{A}\right)^{2}}\right)^{\frac{1}{2}}
\,\frac{4\,\sqrt{\nu\left(1-\nu\right)}}{u\left(\nu\right)}
\label{Rcgendrop}
\end{equation}
 for the drops and
\begin{equation}
R_{c}\,=\,\left(\frac{\sigma\,\varepsilon_{0}}{e^{2}\,
\left(n_{B}-n_{A}\right)^{2}}\right)^{\frac{1}{2}}\,\frac{2}{u\left(\nu\right)}
\label{Rcgenstripes} 
\end{equation}
for the stripes.

Notice that the exponent could be anticipated from dimensional analysis. For arbitrary
dimensionality $d$ of the system we have:
$$R_{c}\sim \left(\frac{\sigma\,\varepsilon_{0}}{e^{2}\,
\left(n_{B}-n_{A}\right)^{2}}\right)^{\frac{1}{d}}$$ times a function of $\nu$.

Once $ R_{c} $ has been eliminated the mixing energy can be put in a
common expression together with the 3D case of
Ref.~\onlinecite{lor01I,lor01II,lor02} as: 
\begin{equation}
e_{m}\,=\,\left(\frac{\sigma^{d-1}e^{2}
\left(n_{B}-n_{A}\right)^{2}}{\varepsilon_{0}}\right)^{\frac{1}{d}}\,u\left(\nu\right)
\label{mixing}
\end{equation}
where:
 \begin{equation}\label{eq:udrop}
u\left(\nu\right)\,=\,\frac{8}{\sqrt{3}}\,\nu\left(1-\nu\right) 
\end{equation}
for 2D drops, while:
\begin{equation}\label{eq:ustripe}
u\left(\nu\right)\,=\,2\sqrt{2}\,\nu\left(1-\nu\right)\left[-\log \nu
\left(1-\nu\right)\right]^{\frac{1}{2}} 
\end{equation}
for 2D stripes. The corresponding functions in the
3D case can be found in Ref.~\onlinecite{lor01I}. The function $u\left(\nu\right)$ encodes all the information about the
geometry and it is represented in Fig.~\ref{ucomparison} for the 2D
case.

We see from Fig.~\ref{ucomparison} that only close to $\nu\sim0$ 
and $\nu\sim1$ the two geometries compete,
since for intermediate volume fractions the stripe geometry is stabilized. The interplay 
among the two states is therefore well described by the analytical
approximation for the electrostatic energy 
Eqs.~\eqref{electrodrops},\eqref{electrostripes}. Comparing the two
curves for the stripe geometry we see that the 
use of the exact numerical expression will only produce small changes 
in physical quantities for $\nu \sim 1/2$
but cannot change the qualitative behavior. Thus, our use of the  
analytical expressions
(Eqs.~\eqref{electrodrops},\eqref{electrostripes}) of $e_{el}$ is justified.

 We minimize the total free energy per unit volume with respect to
$n_{B}-n_{A}$ and $\nu$ as in Ref.~\onlinecite{lor01I} and obtain the coexistence equations for both 2D and 3D systems:
\begin{equation}
\begin{array}{l}
\mu_{B}\,-\,\mu_{A}=-\dfrac{2}{d}\left(\dfrac{\sigma^{d-1}\,e^{2}}
{\varepsilon_{0}}\right)^{\frac{1}{d}}
\dfrac{n_{B}\,-\,n_{A}}
{\left|n_{B}\,-\,n_{A}\right|^{2-\frac2d}}
\dfrac{u\left(\nu\right)}{\nu\,\left(1-\,\nu\right)}
\\ \\ \\
\begin{array}{lll}
p_{B}-p_{A} &=&
\,\left[n+\left(n_{B}-n_{A}\right)\,\left(1-2\,\nu\right)\right]\,\left(\mu_{B}\,-\,\mu_{A}\right)+
\\ \\ & &
\left(\dfrac{\sigma^{d-1}\,e^{2}\,\left(n_{B}\,-\,n_{A}\right)^{2}}
{\varepsilon_{0}}\right)^{\frac{1}{d}}\,\dfrac{\partial{u\left(\nu\right)}}{\partial{\nu}}
\end{array}
\end{array}
\label{coexeqn}
\end{equation}
where $$\mu_{A,B}=\dfrac{\partial f_{A,B}}{\partial n_{A,B}}$$ and the
pressure is defined as:
\begin{equation}
  \label{eq:defp}
p_{A,B}=-f_{A,B}+\mu_{A,B}\,n_{A,B}  
\end{equation}
 We have neglected a possible
dependence of $ \sigma $ on density which contributes with additional terms
(see Ref.~\onlinecite{lor01I}).

Eq.~\eqref{coexeqn}  generalizes the ordinary Maxwell construction to
the charged system case. The former ($\mu_{A}\equiv\mu_{B}$ and
$p_{A}\equiv p_{B}$) is recovered when
$e\rightarrow0 $. 
The first of Eq.~\eqref{coexeqn} states that the most dense
phase has the lowest chemical potential.
The difference of chemical potentials becomes particularly simple for
 2D drops. In this case it  
is (within our approximations) constant and equal to 
$\pm  8 \sigma^{1/2} e/(3 \varepsilon_{0})^{1/2}$.

\section{Separation between a compressible and an incompressible
  phase}
\label{sec:separ-betw-compr}

Now we want to investigate an inhomogeneous state for a 2D system  which involves an incompressible A-phase hereafter ``the
insulator'' and a compressible B-phase hereafter ``the
metal''.\cite{note}  

This will be appropriate for example for a doped Mott insulator.
The case of the 2DEG requires consideration of phases with negative
compressibility and its full discussion will be presented elsewhere.\cite{ort06}

Since the insulator is incompressible and thus electrostatically  
inactive, we have to consider only the excess density of mobile 
electrons counting from the insulating state. Without loss of 
generality we set $n_{A}\equiv 0$ and 
 $f_{A}\equiv 0$ for the insulator. In this way our density $n$ has
 the meaning of density deviation from the pure insulating phase. 
Also without loss of generality we consider $n>0$ so that the
 insulating phase is at low density and the metallic phase is at high
 density but our results apply equally well to the opposite case
 simply by changing $n\rightarrow -n$. With these conventions 
we are reduced to study the problem of the
phase-separation between a metal at finite $n_{B}$ and ``void'' (playing the
 role of the insulator). Both phases are 
 in the presence of a uniform background. 
In the present case the neutrality constraint reduce to: 
 $ \nu\,n_{B}=n $, while
minimizing the free energy in the phase separated state the two 
Eqs.~\eqref{coexeqn} reduce to a single equation for the B-phase pressure:
\begin{equation}
  \label{eq:pbpm}
p_{B}(n_B)=p_{m}
\end{equation}
where the function $p_{B}(n_B)$ is a property of the bulk phase 
[c.f. Eq.~\eqref{eq:defp}] and
\begin{equation*}
p_{m}\,=\,\left(\frac{\sigma\,e^{2}}{\varepsilon_{0}}\right)^{\frac{1}{2}}n_{B}\left[\frac{\partial{u\left(\nu\right)}}{\partial{\nu}}
-\frac{u\left(\nu\right)}{\nu}\right]
\end{equation*}
Eq.~\eqref{eq:pbpm} with the neutrality constraint determines the
behavior of the local density and volume fraction as a function of $n$.

In the limit of $ e\rightarrow0 $ we obtain the equation 
$p_{B}(n_B)=0 $ which corresponds to Maxwell construction for the case
of phase separation between a self-bound neutral fluid and vacuum. 

 When $e\neq0$, $p_{m}$ represents the pressure
due to the presence of the long-range Coulomb interaction and the
surface energy, that we label ''mixing pressure'' and must be
balanced by the B-phase pressure. 

For the drop geometry one has:
\begin{equation}
p_{m}\,=\,-\frac{8}{\sqrt{3}}\left(\frac{\sigma\,e^{2}}{\varepsilon_{0}}\right)^{\frac{1}{2}}\,n
\label{pmixdrop}
\end{equation}
which depends on the global density of the system only. 

Instead, in the stripes geometry the mixing pressure depends explicitly on the local density of the inhomogeneities:
\begin{eqnarray}
p_{m}\,&=&\,\frac{\left(\dfrac{\sigma\,e^{2}}{\varepsilon_{0}}\right)^{\frac{1}{2}}}{\left[-\ln\dfrac{n}{n_{B}}\left(1-\dfrac{n}{n_{B}}\right)\right]^{\frac{1}{2}}}
\left[2\sqrt{2}\,n \right.  + \nonumber \\ 
\label{pmixstripe}
\\ & &
\left. 2\sqrt{2}\,n\,\ln\left[\frac{n}{n_{B}}\left(1-\frac{n}{n_{B}}\right)\right]\,-\sqrt{2}\,n_{B}\right]
\nonumber
\end{eqnarray}
 The mixing pressure is negative for both geometries at all volume
fractions. This means that the metallic phase is under ``tensile
stress'' due to the long-range interaction. Thus the equilibrium density of the metal
is lower than the density predicted by MC in the neutral case. For a discussion on the stability of a
fractionated electronic fluid subject to a negative pressure see 
the Appendix B of Ref.~\onlinecite{lor01I}.

From the definition of the
pressure one finds for the B-phase chemical potential:
\begin{equation*}
\mu_{B}=\frac{f_{B}\left(n_{B}\right)}{n_{B}}+\frac{p_{m}}{n_{B}}
\end{equation*}
The last term is the contribution due to the frustrating forces. 
For the global chemical potential, 
$\mu=\partial f/\partial n$, we find:
$$
\mu=\frac{f_{B}\left(n_{B}\right)}{n_{B}}+\mu^{mix}
$$
with the mixing contribution given by:
$$
\mu^{mix}=\frac{\partial n_{B}}{\partial
  n}\,\frac{n}{n_{B}^{2}}\,p_{m}  
+\frac{d}{d n} e_{m}[n_{B}(n),\nu(n)] \nonumber.
$$

We have provided the general equations for a FPS between an incompressible phase at $n_{A}=0$ and a compressible phase.
To proceed further we need an explicit expression for the bulk free energy.
\subsection{Parabolic approximation for 2D metal free energy}
\label{secpstrans}
Now we solve our problem expanding the free energy of the B-phase in a parabolic approximation around the density at which, in absence of the long-range Coulomb interaction, the system would
experience the transition to the
phase-separated state. In other words we will label $ n_{B}^{0} $ the
density which satisfies the equation $ p_{B}\left(n_{B}\right)=0 $,
$f_{B}^{0}$ the corresponding free energy
$f_{B}^{0}\equiv f_{B}\left(n_{B}^{0}\right)$ and $\mu_{0}$ the corresponding chemical
potential. These quantities are related by $f_{B}^{0}=\mu_{0} n_{B}^{0}$. 

 Using a Taylor expansion we can then write the B-phase free energy as follows:
\begin{equation}
f_{B}\left(n_{B}\right)\,=\,
\mu_{0}\,n_{B}+\,\frac{1}{2\,\kappa_{B}}\left(n_{B}-n_{B}^{0}\right)^{2}
\label{eq:paraener}
\end{equation}
where $\kappa_{B}=\left(\partial^{2}f/\partial^{2} n\right)^{-1} $ is proportional to the compressibility of the metallic
phase. 

The pressure of the B-phase is now:
\begin{equation}
p_{B}\left(n_{B}\right)\,=\,\frac{1}{2\,\kappa_{B}}\left(n_{B}^{2}-n_{B}^{0^{\,2}}\right)
\label{parapressure}
\end{equation}

\begin{figure}[tbp]
\includegraphics[width=8cm]{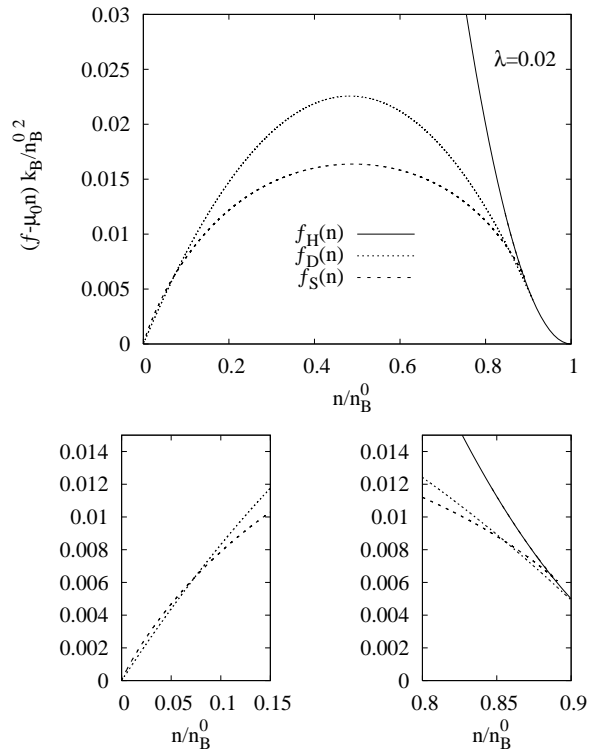}
\caption{Top: The free energy of the uniform metallic phase
  $f_{H}\left(n\right)$ compared with the FPS-state
  in the case of drops  [$f_{D}\left(n\right)$] and stripes
  [$f_{S}\left(n\right)$] for  $\lambda=0.02$.  
 Bottom: Expanded scale in the density range where the geometries
  compete.  }
\label{fconlow}
\end{figure}

We introduce a dimensionless parameter $\lambda$ that
measures the strength of the frustration due to the mixing energy
effect. $\lambda$ can be defined as the ratio of the characteristic 
mixing energy (obtained from Eq.~\eqref{mixing} without the geometric factor $u(\nu)$) 
to the characteristic phase separation energy gain $\sim n_{B}^{0\,2}/\kappa_{B}$:
\begin{equation*}
\lambda =\frac{\kappa_{B}}{(n_{B}^{0})^2}\,\left[\frac{\sigma^{d-1} e^{2}(n_{B}^{0})^2}{\varepsilon_{0}}\right]^{\frac{1}{d}}
\end{equation*}
 $\lambda$ coincides apart from a numerical factor with the parameter
introduced in Ref.~\onlinecite{lor01I} for $d=3$. 
In the following we measure the pressure and free energy densities in units of
the characteristic phase-separation energy gain $(n_0^{B})^2/k_B$
and  all the densities in units of the characteristic MC density
$n_{B}^{0} $. Thus we define 
$n_{B}^{\prime}=n_{B}/n_{B}^{0}$,
$n^{\prime}=n/n_{B}^{0}$, $p_{B}^{\prime}=p_{B}\kappa_{B}/
n_{B}^{0^{\,2}}$, the uniform $B$ phase energy density 
$f_H=f_{B}{\kappa_B}/{n_0^{B\;2}}$ and the phase separated energy density
$f_{D,S}=f{\kappa_B}/{n_0^{B\;2}}$.
In our parabolic approximation the free energy
densities read:
\begin{eqnarray}
f_{H}\left(n^{\prime}\right)&=&\mu_{0}\,n^{\prime}\,\frac{\kappa_{B}}{n_{B}^{0}}\,+\,\frac{1}{2}\left(n^{\prime}-1\right)^{2}
 \label{fpsfreeene}
\\ \nonumber \\
f_{D,S}\left(n^{\prime}\,,n_{B}^{\prime}\right)&=&\mu_{0}\,n^{\prime}\,\frac{\kappa_{B}}{n_{B}^{0}}\,+\,\frac{n^{\prime}}{n_{B}^{\prime}}\,\frac{1}{2}\left(n_{B}^{\prime}-1\right)^{2}+ \nonumber
\\ & &
\lambda\,n_{B}^{\prime}\,u\left(\frac{n^{\prime}}{n_{B}^{\prime}}\right) \nonumber
\end{eqnarray}
The local density $n'_B$
is determined by solving Eq.~\eqref{eq:pbpm} with the left and right
hand side given by Eq.~\eqref{parapressure} and
Eqs.~\eqref{pmixdrop} and ~\eqref{pmixstripe} respectively.

To decide the most stable geometry for the FPS and its stability
against the uniform phase 
we have to
compare the expressions in Eqs.~\eqref{fpsfreeene} for different
strengths of the LRC interaction.
The first term in the free energies
represents the MC free energy which is
equal for all states and can be eliminated.

\begin{figure}[tbp]
\includegraphics[width=8cm]{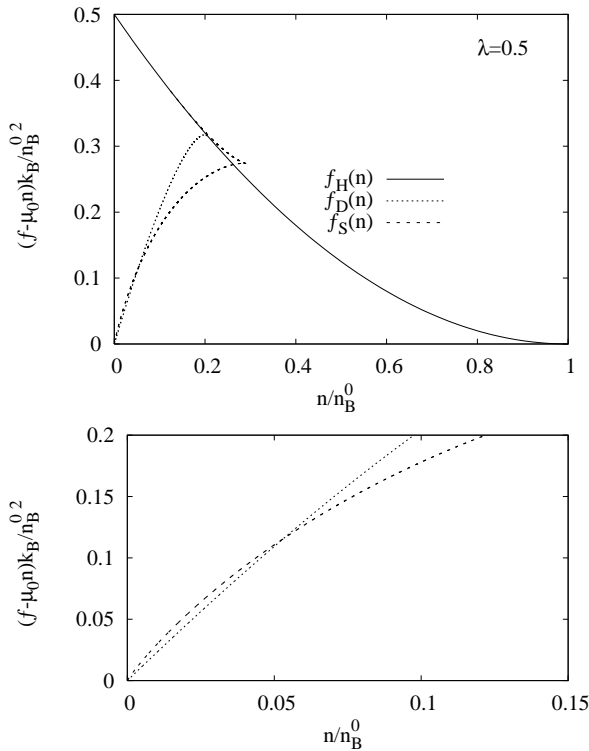}
\caption{Same as Fig.~\ref{fconlow} for $\lambda=0.5$.}
\label{fconhigh}
\end{figure} 
 
 In Figs.~\ref{fconlow},\ref{fconhigh} we plot the free
 energies with this term subtracted. We show the free energies 
of the uniform and
the FPS-state in the presence of drops ($f_D$) and stripes ($f_S$) 
for two different values of the parameter $\lambda$.
 For small $\lambda$ (Fig.~\ref{fconlow})
  the striped state introduces
  a lower energetic cost with respect to the drop geometry at intermediate global densities, while at low and
  high densities the two geometries compete. Increasing the
  global density one finds a first ``geometrical''
  transition from a droplet state to a striped one at low density (left bottom panel) and a second ``geometrical''
  transition from stripes to drops at high density 
(right bottom panel).
 For large $\lambda$ the situation is different 
(Fig.~\ref{fconhigh}).The striped state
results stable for both intermediate and high density and the
two geometries compete only at low density (bottom panel) where
there is a ``geometrical'' transition from the drops to the stripes as
in the previous case.

Note that the free energies in the FPS-state
for both geometries have a negative curvature and thus exhibit
a negative electronic compressibility. 
This, however, does not imply necessarily
an instability since the background
contribution to the inverse compressibility, which is very large and 
positive\cite{lor01I}, should also be taken into account.

\begin{figure}[tbp]
\includegraphics[width=8cm]{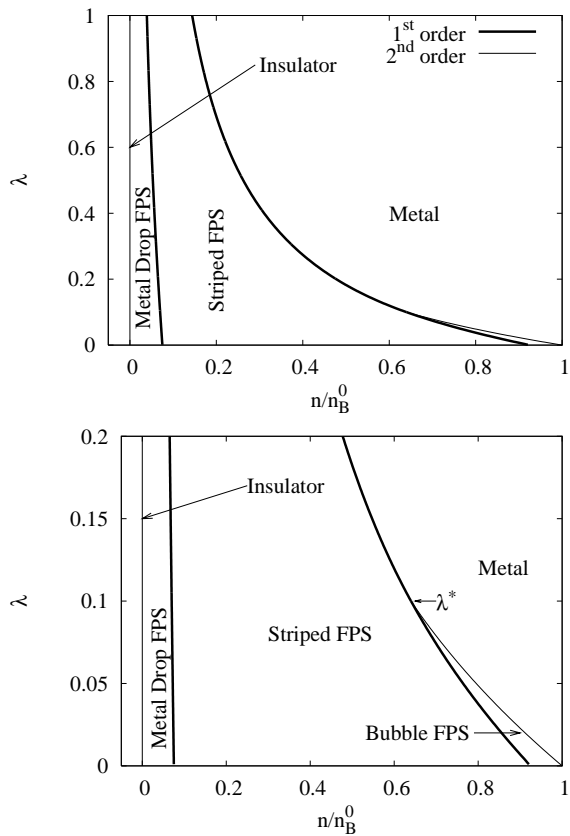}
\caption{Top: The phase-diagram in the $n-\lambda$ plane. Between the two uniform
  phases one finds the FPS-states. The
  transition from the insulator to the metal drop state is second order at
  all $\lambda$ ({\it see} Sec.~\ref{secpsprop}), while the transition
  between the metal and the striped FPS-state at
  $\lambda>\lambda^{\star}$ is first order as well as the
  ``geometrical'' transitions. Bottom: Expanded scale for $\lambda \simeq
  \lambda^{\star}$. One finds a second-order transition from the metal to the ``bubble'' state.  }
\label{pd}
\end{figure}

 The range of global density where the FPS-state
is stable depends upon $\lambda$.
Furthermore inside the FPS-state one finds that the stable geometric
arrangement changes both with the global density and the strength of
the LRC interaction. 
This leads to construct a phase-diagram in the $n-\lambda$ plane (see Fig.~\ref{pd}).
 Given our initial choice we have two uniform phases, the metallic
 phase at high global densities and the insulating phase at
 $n=0$. 
 For all $\lambda$ the two uniform phases are  
separated by FPS-states. The global density range of stability of
the FPS, which for the unfrustrated case in absence of the LRC interaction is determined by $0<n^{\prime}<1$ ({\it i.e.} $0<n< n_{B}^{0}$),
shrinks increasing $\lambda$.
This is clearly due to the 
tendency of the long-range Coulomb interaction to stabilize the
uniform state as in 3D systems\cite{lor01I}. 
Close to the insulator one finds metal drops and close to the uniform
metal one finds circular voids in the metal. We will call the latter
``the bubble state''. Finally at intermediate densities one finds stripes.
Increasing $\lambda$, the bubble state stability range shrinks and
disappears above a value $\lambda^*\sim 0.1$ so that entering from
the metallic uniform phase the FPS-state is made of stripe
inhomogeneities. On the contrary the low density
metallic drop state persists at all $\lambda$.  

\subsubsection{Properties of the frustrated phase separated state}
\label{secpsprop}
We now discuss explicitly the physical properties of the FPS-state and
the order of the various transitions.
For the stripe state the relation for the local
density as a function of global density is 
obtained in terms of $\nu$ in a parametric form.
Then the 
volume fraction is obtained from the neutrality constraint. 
When the inhomogeneous phase has a drop geometrical arrangement (metal
drop state and bubble state) the local density in the metal takes a
particularly simple form.  Solving 
Eqs.~\eqref{eq:pbpm},\eqref{pmixdrop},(\ref{parapressure}) with respect to $n_{B}$ we derive
the B-phase local density in
term of the global density:
\begin{equation}
n_{B}^{\prime}\,=\,\sqrt{1-\frac{16}{\sqrt{3}}\,\lambda\,n^{\prime}}
\label{localdensitydrops}
\end{equation}
Obviously in the case $\lambda= 0$ one recovers the MC solution $n_{B}=
n_{B}^{0}$.  
Contrary to the MC 
the local density of the metallic inhomogeneities depends
explicitly on
and is a
decreasing function of the global density $n$ of the system. Increasing the strength of the long-range
interaction $\lambda$, this effect becomes stronger. 

This behavior of the local density versus global density 
can be detected from local probes (like NQR and NMR).
Physical quantities, which depend on the local density, will show an
unexpected behavior: they will respond to an increase in density
as if the density were decreasing. This effect has been discussed 
in connection with the Curie temperature  in manganites.\cite{lor01II,lor02pb}

\begin{figure}[tbp]
\includegraphics[width=8cm]{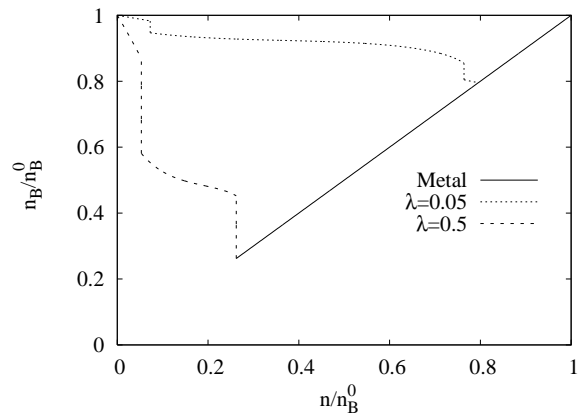}
\caption{The local density behavior vs. the global density for
  $\lambda=0.05, 0.5$. In correspondence to the ``geometrical''
  transitions and at the transition from the metal to the striped
  state 
 $n_{B}$ shows a discontinuity reflecting their first-order
  nature. The segments close to the origin corresponds to metallic
  drops. The small segment in the upper curve close to the full line
  is the bubble state. The rest of the dashed curves correspond to
  stripes. The full line is the uniform metal.}
\label{nblocallambda}
\end{figure}

\begin{figure}[tbp]
\includegraphics[width=8cm]{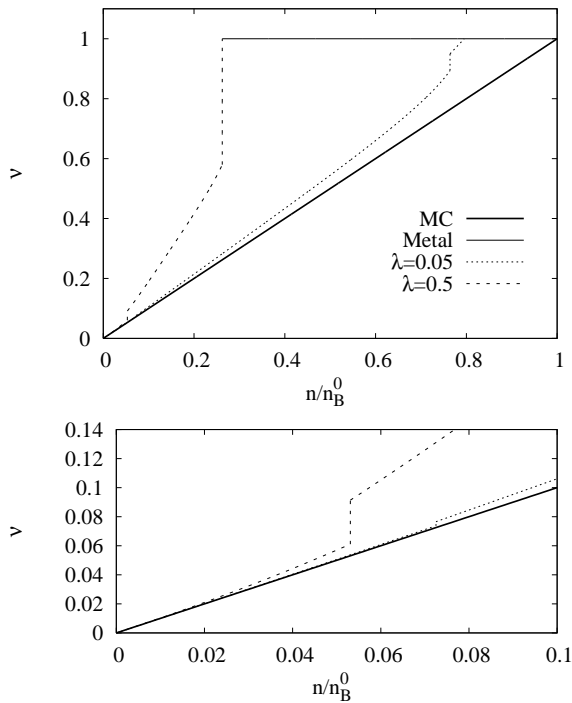}
\caption{Top: The volume fraction of the B-metallic phase in the FPS-state
  vs $n$ 
at different values of $\lambda$ compared with the ordinary Maxwell
  construction  analysis. The full horizontal line corresponds to the uniform
  metal. For $\lambda=0.05$ the small segment close to the uniform
  metal is the bubble state. Close to the origin for both values of
  $\lambda$ one has the drop
  state. The rest is in the stripe state.  
Bottom: Expanded scale at low density to show the
  abrupt change of $\nu$ due to the ``geometrical'' transition from
  the metal drop state to the striped state as density increases.}
\label{nulambda}
\end{figure}

For the metal drop state at any $\lambda$, $n_{B}$ approaches the
MC value $ n_{B}^{0} $ going towards the homogeneous insulating phase,
{\it i.e.} $n \rightarrow 0$ (Fig.~\ref{nblocallambda}).
Furthermore the transition from the FPS to the
insulating-phase is second order like, 
since the volume fraction goes continuously to zero with 
sloop 1 ($\nu\sim n'$) (Fig.~\ref{nulambda}).
From the same figure we see that for 
$\lambda=0.05<\lambda^{\star}$, the system goes from the uniform metal to the bubble state
with the volume
fraction for the insulating phase (given by $1-\nu$) which goes continuously to
zero at the transition. 
Also the local density of the metallic regions
is continuous (upper curve in Fig.~\ref{nblocallambda})  
indicating that the transition is second-order like. 

For $\lambda>\lambda^{\star}$ the 
region of stability of the bubbles close to $\nu\simeq 1$ 
disappears and the striped FPS-state appears at the
transition with insulating inhomogeneities that have a finite volume fraction (Fig.~\ref{nulambda}).  
From these properties one can conclude that the transition is
first-order in this case. The first order character of the transition can be understood 
from the behavior of the mixing pressure for the stripe case.
From Eq.~\eqref{pmixstripe} one notes that in this case the mixing
pressure has a divergence at $n_{B}=n$ which cannot be
reached. Therefore one can not go continuously from the uniform metal
to the stripe state.
 
Inside the FPS-state one can see that the high density 
``geometrical'' transition
from the ``bubble'' state to the striped one for
$\lambda<\lambda^{\star}$ and the low density ``geometrical''
transition from the latter state to the metal drop state have a
discontinuity of both the volume fraction and the B local density
[Figs.~(\ref{nblocallambda}),(\ref{nulambda})] 
reflecting the first-order nature of the transitions.
This is expected since at least in our approach one can not
continuously deform drops to get stripes.
 
For all the FPS-states, increasing the
strength of the long-range interaction the metallic density decreases 
(Fig.~\ref{nblocallambda})
in order to minimize the mixing energy of the inhomogeneous state which is
$\propto\lambda\,n_{B}$. 
At the same time, 
increasing $\lambda$ (Fig.~\ref{nulambda}),  
the volume fraction has a growing
rate larger than in the ordinary MC case. 

\begin{figure}[tbp]
\includegraphics[width=8cm]{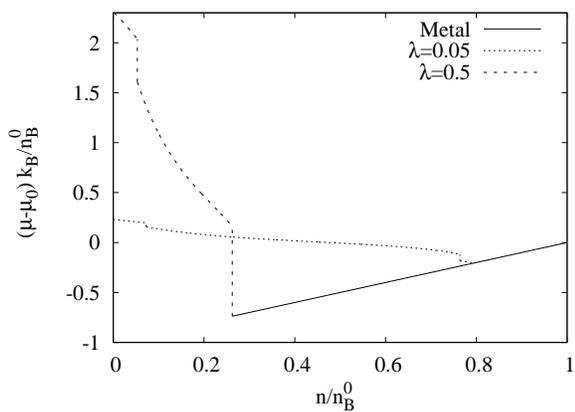}
\caption{The chemical potential behavior in the FPS-state. When the
  ``bubble'' state exists, at the transition from the
  metal to the FPS-state $\mu$ exhibits a cusp point ($2^{nd}$ order
  transition) while for $\lambda>\lambda^{\star}$ the transition to
  the striped state implies a chemical potential discontinuity
  ($1^{st}$ order transition)}
\label{mulambda}
\end{figure}

The order of the  transitions can be also checked 
if one looks to the chemical potential
 $\mu=\partial f/\partial n$ (Fig.~\ref{mulambda}). 
In the range of stability of the FPS ``bubble'' state (small $\lambda$) the
chemical potential at the critical density has a cusp indicating the
second-order nature of the transition. 
Increasing $\lambda$ to values greater than $\lambda^{\star}$ the
``bubble'' state disappears and the chemical potential has a
discontinuity at the transition to the striped FPS-state.
A similar discontinuity in $\mu$ is also obtained in correspondence to
the ``geometrical'' transitions consistent with the fact that these transitions
are first-order like. Notice that for the insulator, which is
 incompressible, the chemical potential is not defined.  

Finally we analyze the size $ R_{d} $ of the inhomogeneities 
and the size $R_{c}$ of the cells in which the system is divided. 
One can introduce a length scale $l_{d}$ which fixes the natural units 
for the characteristic size of the cell. This scale correspond to $
R_{c} $ (Eqs.~\eqref{Rcgendrop},~\eqref{Rcgenstripes}) 
evaluated at the MC-density $n_{B}^{0}$ dropping all the geometric
factors:
\begin{equation}
l_{d}\,=\,\left(\frac{\sigma\,\varepsilon_{0}}{e^{2}(n_{B}^{0})^2}\right)^{\frac{1}{2}}
\end{equation}
Another important length scale is the screening length: 
\begin{equation}
l_{s}=\frac{\varepsilon_{0}}{2 \pi e^{2}\,\kappa_{B}}
\end{equation} 
Inserting the last equation in the definition of $\lambda$ one finds
 that $\lambda=l_{d}/(2\pi l_{s})$.
Similar length scales where defined in the 3D
 case,\cite{lor01I} in particular we can write the 
screening length for general dimension (neglecting factors of  $\pi$) 
 $l_{s}\sim (\varepsilon_{0}/(e^{2}\kappa_{B}))^{1/(d-1)}$.
In $3D$ systems $l_{S}$ gives an upper limit for the size of inhomogeneities.\cite{lor01I,lor01II,lor02} 
The nature of screening is quite 
different in 2D systems.\cite{and82} 
As it will be clear in Sec.~\ref{seclda}, $l_s$ still plays a 
fundamental but does not limit the size of inhomogeneities. 
\begin{figure}[tbp]
\includegraphics[width=8cm]{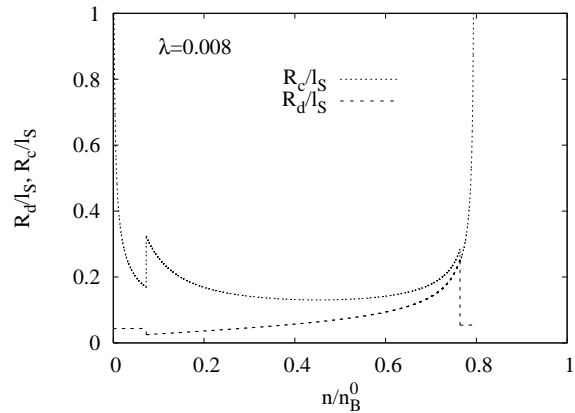}
\caption{The cell radius and the inhomogeneity radius vs the global
  density for small $\lambda$. The divergence in $R_{c}$ are in correspondence to the
  transitions from the FPS-state to both the metal and the
  insulator.}
\label{Rnlow}
\end{figure}

\begin{figure}[tbp]
\includegraphics[width=8cm]{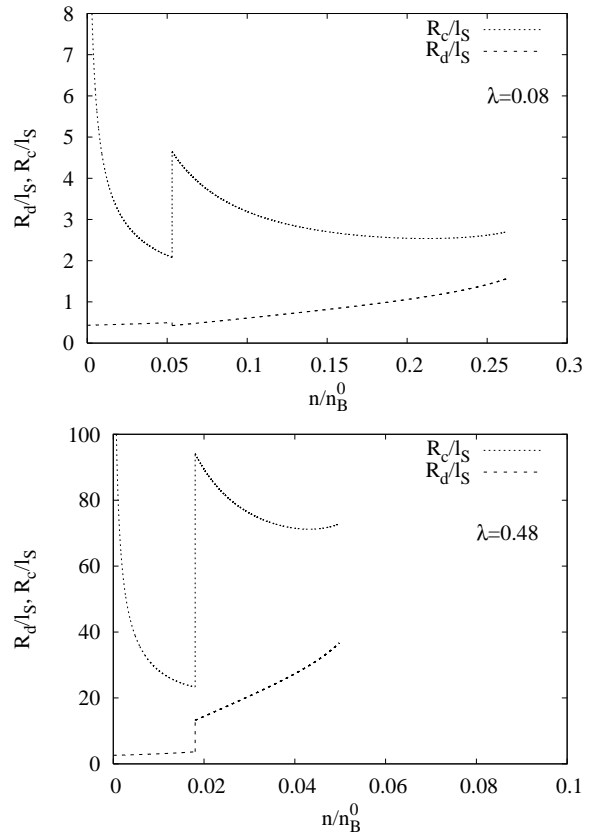}
\caption{Top: $R_{c}$ and $R_{d}$ vs. $n$ for $\lambda$=0.08. The FPS-state appear with a finite cell radius. The
  typical size of the inhomogeneities is of the order of $l_{S}$. Bottom: Same of top for $\lambda=0.48$. The sizes of the domains
  and the cells are much greater than $l_{S}$. }
\label{Rnhigh}
\end{figure}

With the above definitions  one has:
\begin{equation}
R_{c}\,=\,2 \pi\lambda\,l_{s}\,\frac{1}{n_{B}'}\,\frac{2}{u\left(\nu\right)}\,\,\,\,\,\,\,\, \text{Stripes}
\end{equation}
\begin{equation}
R_{c}\,=\,2 \pi \lambda\,l_{s}\,\frac{1}{n_{B}'}\,\frac{4\,\sqrt{\nu\left(1-\nu\right)}}{u\left(\nu\right)}\,\,\,\,\,\,\,\, \text{Drops}
\end{equation}
For the stripes the half-width of the metallic regions is given 
by $R_{d}=R_{c}\nu$. For the drop geometry it is more convenient to 
define a symmetrized radius $R_{d}=R_{c}[\nu(1-\nu)]^{\frac{1}{2}}$ that smoothly interpolates between
the  radius of metallic drops at small $\nu$  and the radius of
bubbles for $\nu$ close to 1. The above assumptions  
 lead to the simple
expression,
\begin{equation}
  \label{eq:rddrop}
  R_d= \sqrt3 \pi \frac{\lambda l_s}{\sqrt{1-\frac{16}{\sqrt{3}}
      \lambda n'}}= 
\frac{\sqrt3}2 \frac{l_d}{\sqrt{1-\frac{16}{\sqrt{3}}\lambda n'}}.
\end{equation}
For small $n'$ or small $\lambda$ one has an almost constant behavior $R_d\sim0.9 l_d $ 
[c.f. Figs.~\ref{Rnlow},\ref{Rnhigh}].

  For $\lambda<\lambda^*$ bubbles appear in the metal with a divergence of the cell radius
 while the drop radius remains finite  ({\it see} Fig.~\ref{Rnlow}). 
That is at threshold bubbles appear suddenly with a finite size 
but the transition is second order because they are infinitely far
apart.
 This is reminiscent of the transition in a type II superconductor as
 a function of field at $H_{c\,_{1}}$, which according to GL theory is
 second-order
although normal state ``drops'' (the vortex core) have a finite radius $\xi$.
\cite{tin75}

 For $\lambda<\lambda^{\star}$, $R_{d}$ is of the order of $l_{s}$.
 If the latter is of the order of the interparticle distance, a mesoscopic
 treatment may be problematic. 
On the other hand increasing $\lambda$,
$R_d$ grows to values that are much larger than $l_{s}$ (Fig.\ref{Rnhigh}). This
represents the main difference respect to 3D systems for
which it was demonstrated that $l_{s}$ is an upper bound for $R_{d}$ at
any $\lambda$. In the next section we will show that this difference
stands on the different role of the LRC interaction and screening
in 2D and 3D systems. 

The divergence of the cell radius 
disappear when $\lambda$ is greater than $\lambda^{\star}$, since at
the first-order transition to the striped state the cell radius stays
finite.
 
Increasing $\lambda$ at fixed global density one finds that $R_{c}/l_s$
increases. This behavior is easy to rationalize if one considers an
increase of the surface energy in such a way that $\lambda$ increases while keeping  
$l_s$ constant. In this case the system prefers to make domains
with longer periodicity to reduce the surface energy. From Fig.~\ref{Rnhigh} we see that metallic stripes become narrower as
the insulator is approached.

\section{Local density approximation}
\label{seclda}
In this section we want to relax the uniform density approximation.
For simplicity as in Sec.~\ref{sec:separ-betw-compr} we restrict ourselves 
to study a phase separation between a metallic
phase $B$ and an insulating phase $A$ with energy density $f_A=0$.

In order to account for the
spatial dependence of the local density, we take the total
free energy as a local density functional.   
Indicating with
$V_{B}$ the B-phase domain volume and with $V$ the total
system volume the total free energy can be put as:
\begin{eqnarray}
 F& =& \int_{\bm{r}\,\epsilon\, B} f_{B}
\left(n_{B}\left(\bm{r}\right)\right) d^{\,2}r
+\sigma \Sigma_{A\,B} +\frac{1}{2\,\varepsilon_{0}}\,\int\!\!\!\!\int
\dfrac{e^{2}}
{\left|\bm{r}_{1}-\bm{r}_{2}\right|}
\nonumber\\&&\left[n_{B}\left(\bm{r}_{1}\right)-
\overline{n}\right]
\left[n_{B}\left(\bm{r}_{2}\right)-\overline n \right] d^{\,2}r_{1}
d^{\,2}{r_{2}} \label{enerlda}
\end{eqnarray}
where $\overline{n}$ is the global density, while
$\Sigma_{A\, B}$ is the total interface surface. We still assume a 
sharp interface due to short range forces with the parameter 
$\sigma$ parameterizing the surface energy.  Eventually the surface
energy term could be replaced by a gradient term to consider soft interfaces.

The constraint of charge neutrality of the system reads:
\begin{equation}
\int_{\bm{r}\,\epsilon\, B}\,n_{B}\left(\bm{r}\right) d^{\,2}r=\overline{n}\,V
\label{ldavincoloneutralita}
\end{equation}
 We minimize the total free energy functional with respect to the
local density and include the above neutrality constraint 
via a  Lagrange multiplier $\mu_{e}$.
We then obtain an equation which tell us that the 
electrochemical potential is constant:
\begin{equation}
\mu_{B}\left(\bm{r}
\right)-\frac{e}{\varepsilon_{0}}\phi\left(\bm{r}\right)=\mu_{e}
\quad \forall\, \bm{r}\,\epsilon\,B
\label{potenzialeelettrochimico}
\end{equation}
 where $\phi\left(\bm{r}\right)$ indicates the electrostatic potential
 generated by the charge distribution
 $\left[n_{B}\left(\bm{r}\right)-\overline n\right]$. This equation
 has to be solved together with the neutrality condition 
Eq.~\eqref{ldavincoloneutralita}.

Assuming again for the B-phase the
parabolic free energy [Eq.~\eqref{eq:paraener}],
one obtains an  equation relating the local
density to the potential in terms of the compressibility:
\begin{equation}\label{eq:nbdk}
n_{B}\left(\bm{r}\right)-n_{B}^{0}
           =\kappa_{B}   
           \left[\frac{e}{\varepsilon_{0}}\phi\left(\bm{r}\right)-(\mu_0-\mu_e)\right] 
\quad \forall \, \bm{r}\,\epsilon \, B 
\end{equation}
or in terms of the screening length 
\begin{equation}
\phi\left(\bm{r}\right)-\frac{\varepsilon_{0}}e(\mu_0-\mu_e) =2 \pi e l_s \left[ n_{B}\left(\bm{r}\right)-n_{B}^{0}\right] 
\quad \forall \, \bm{r}\,\epsilon \, B 
\label{minimeq}
\end{equation}
 In the limit of infinite compressibility i.e. zero screening length, the electrostatic
 potential is constant on the metallic regions and  
  therefore $n_{B}\left(\bm{r}\right)$ correspond to the
 distribution of a metal for which the 3D Laplace equation $\nabla^{2}\phi=0$ is
 supplemented by the boundary condition $\phi=\text{{\it const}}$ on
 the domains. 
Here we are assuming that $l_s/R_d \to 0$ but
 we are not making assumptions about the other parameters so that
$\lambda\sim l_d/l_s$ may remain finite.

\begin{figure}[tbp]
\includegraphics[width=8cm]{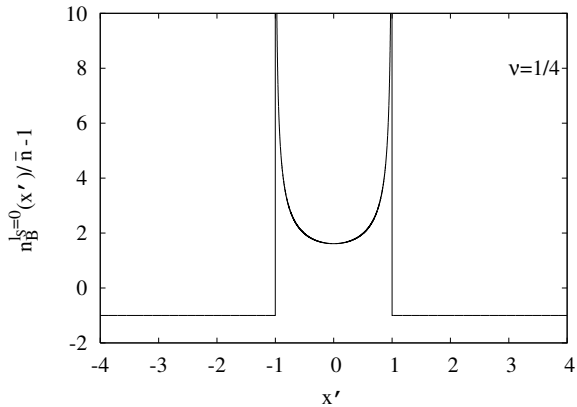}
\caption{The excess charge density profile
  $n_{B}^{l_s=0}\left(x^{\prime}\right)/\overline n -1$ in the direction
  perpendicular to the stripe for $ \nu=1/4 $. $x^{\prime} $ is the component of $\bm{r}/R_{d}$}
\label{perfectmetal}
\end{figure}

 In this limit the problem can be solved analytically for the stripes geometry. 
In fact, in this case the Coulomb potential is calculated by using the 
Schwarz-Christoffel conformal transformations.\cite{smy68} The B-phase density
spatial dependence comes out to be:
\begin{equation} 
n_{B}^{l_{S}=0}\left(x^{\prime} \right)\,=\,\overline{n}\:\dfrac{\left|\mbox{cos}\,\dfrac{\pi
x^{\prime} \nu}{2}\right|} {\sqrt{\mbox{sin}^{2}\,\dfrac{\pi\nu}{2}\,-\,\mbox{sin}^{2}\,\dfrac{\pi x^{\prime}\nu }{2}}}
\label{densitalocale0ordine}
\end{equation}
where $x^{\prime}$ indicates the $x$ component of the dimensionless
coordinate $\bm{r}'$ defined by $\bm{r^{\prime}}=\bm{r}/R_{d}$.
In Fig.~(\ref{perfectmetal}) we show the spatial dependence of the excess charge
density $ n_{B}^{l_s=0}\left(x^{\prime} \right)/\overline{n} -1 $ on one unit cell.

Keeping in mind that the charge in the stripe region is
undercompensated by the background, one finds that at this level of 
approximation the metallic stripe behaves as 
a macroscopic charged metallic strip. The density accumulates on the
border of the stripe and decays  as a power law towards the center.

It is interesting to
compare this configuration with the three dimensional analogue, 
that is metallic layers locally undercompensated by a uniform background. 
In that case the charge is localized at the surface of the layer and 
decays as $\exp(-x/l_s)$.
Although in both cases $l_s$ is the length scale below which
macroscopic electrostatic, i.e $\phi=\text{{\it const}.}$, 
is not any more valid, the solutions for $x>> l_s$ are dramatically
different in the two cases. For 3D metals the local density far from the surface 
is forced to be equal to the background density which in turn is equal
to the global electronic density i.e. $n_{B}= \overline n$.\cite{lor01I}
On the contrary on the present 2D case the local density  
far from the surface  can be quite different from the global density.

 In Ref.~\onlinecite{lor01I,lor02} it was shown that for small $l_{S}/l_d$ in the
 3D case mesoscopic inhomogeneous phase separation is forbidden. In fact  the
system gains phase-separation energy when the local density differs
from the global density value. Instead electrostatic forces $n_{B}=
 \overline n$, except for the microscopic length $l_{S}$, and the system remains uniform.  
 In the 2D case the inhomogeneity is able to gain phase separation 
energy in all the region where $n_{B}\neq\overline n$ which is not limited by a
 microscopic length.

A related issue in 3D is that in general inhomogeneities   
 can not have all linear dimensions larger than $l_s$. 
In the 2D case this does not impose any constraint because the
linear size perpendicular to the layer is from the outset smaller 
than $l_s^{3D}$.  Therefore  inhomogeneities can have an 
unbound size in the plane. 

Notice that this different behavior between 2D and 3D systems is due
to the difference in the charge profile far from the surface. Of
course the divergence of the electronic density at the surface of the
stripe is unphysical and will be cutoff by a microscopic length (see
below) but this does not affect the behavior
of the charge far from the surface that is essential for our
argument. 

\begin{figure}[tbp]
\includegraphics[width=6cm]{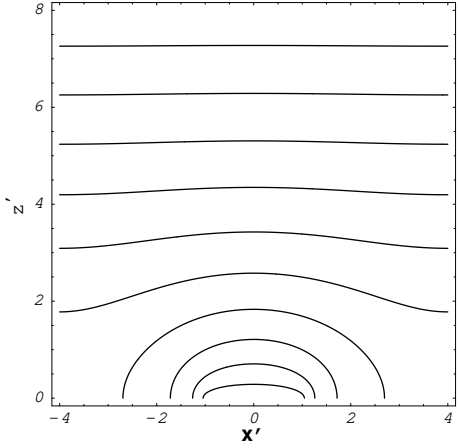}
\caption{The equipotential lines in the plane perpendicular to the
  stripes for the same parameters as Fig.~\protect\ref{perfectmetal}}
\label{potentialperfectmetal}
\end{figure}

In the related problem of a two-dimensional electron (hole) gas in MOSFET
devices the background is not rigid but is provided by a metallic gate
in the form of mobile holes (electrons).  
In this case the background can relax and
other considerations enter into play.\cite{spi03,spi04}
In the case in which the separation between the gate and the e-gas 
is much larger than $R_c$ and $R_d$, the electric field produced by 
the electron gas inhomogeneities will be perpendicular to the gate
plane and this will prevent the background to relax. This is
illustrated in Fig.~\ref{potentialperfectmetal} where we show 
the equipotential lines in the plane perpendicular to the stripe. 
We see that at a distance $\sim R_{c}$ from the stripe the 
 equipotential lines become parallel to the stripe surface. 
In this case the background behaves as rigid and our results
apply. In other words the background can
follow the inhomogeneities of PS  only when the inhomogeneities are comparable in size to the separation between the e-gas and the gate. 
The crossover between these two possible scenarios
will be considered elsewhere.\cite{ort06}.

To further clarify the role of $l_{S}$, now we calculate the charge density profile for
  $l_{S}\neq 0$  adapting the method of Ref.~\onlinecite{and82}. The unphysical divergence
  of the metallic density will be removed.  
We restrict to the stripe geometry as above but the same method can be
  used for other geometries.  Consider the 3D Poisson equation 
 \begin{equation}
   \label{eq:laplace}
  \nabla^{2}\phi^{3D}\left(\bm{r},z\right)=-4\pi \rho\left(\bm{r},z\right)
 \end{equation}
with $z$ the distance from the plane and the 3D charge density defined as
$\rho\left(\bm{r},z\right)=\rho(\bm{r})\delta\left(z\right)$ with
$\rho(\bm{r})$ the 
2D in plane density.
To solve Eq.\eqref{eq:laplace}, we perform the Fourier transform in the $x$ direction using the
fact that the solution is periodic in $2R_c$:
  \begin{eqnarray}
   \phi^{3D}\left(x,z\right)&=&\dfrac{1}{2\,R_{c}}\,\sum_{q}\,e^{iqx}\, \phi^{3D}\left(q,z\right) \,\qquad q=\frac{\pi n}{R_{c}} \nonumber\\
    \phi^{3D}\left(q,z\right)\,&=&\int_{-R_{c}}^{R_{c}} dx e^{-iqx} \phi^{3D}\left(x,z\right)
  \end{eqnarray}   
(and similarly for the charge distribution). 
Since there is no dependence in the direction of the stripe we dropped
the $y$ coordinate.The Poisson equation is then:
  \begin{equation}
  \dfrac{\partial^{2}\phi^{3D}\left(q,z\right)}{\partial
  z^{2}}-q^{2} \phi^{3D}\left(q,z\right)=-4\pi \rho(q)\delta\left(z\right)
  \label{poissonfourier}
  \end{equation}
with $\rho(q)$ defined as the Fourier transform of the in plane density
 $$
  \rho(\bm{r})\equiv -e\left[
  n_{B}\left(\bm{r}\right)-\overline n\right].$$
 For $z\ne 0$ the solution for the potential is
$\phi^{3D}\left(q,z\right)=\phi^{3D}(q,0) e^{-|q||z|}$. Integrating Eq.~\eqref{poissonfourier} 
in a small interval one obtains the boundary condition:
  \begin{equation}
    \label{eq:laplace2d}
    |q| \phi(q)=2\pi \rho(q)
  \end{equation}
with the Fourier transform of the in-plane electrostatic Coulomb potential
defined as
$\phi(q)\equiv\phi^{3D}\left(q,0\right)$. The 3D boundary condition 
 Eq.~\eqref{eq:laplace2d} looks formally like an effective 2D 
Poisson equation.

Eq.~\eqref{eq:nbdk} or Eq.~\eqref{minimeq} determines how the charge responds to the
potential and Eq.~\eqref{eq:laplace2d} determines how the potential 
is generated by the charges. Both equations must be solved
self-consistently in order to find the charge distribution.

Using the superposition principle both charge and
potential can be written as the sum of the terms evaluated above for infinite compressibility ({\it i.e.} for
$l_s=0$) plus a correction, which we 
wish to compute:
  \begin{eqnarray}
 \phi\left(\bm{r}\right)&=&\phi^{l_{S}=0}\left(\bm{r}\right)+\delta\phi\left(\bm{r}\right) \\
 n_{B}\left(\bm{r}\right)&=&n_{B}^{l_{S}=0}\left(\bm{r}\right)+ \delta n_{B}\left(\bm{r}\right)
 \label{splitting}
 \end{eqnarray}

The correction $\delta \phi \left(\bm{r}\right)$ satisfy the effective Poisson equation:
 
 \begin{equation}
 |q|\delta\phi\left(q\right)= - 2\pi e \delta n_{B}(q) \label{eq:lap2d}
 \end{equation}
 
The unknown Lagrange multiplier  $\mu_e$ has to be determined 
by fulfilling the neutrality condition and can also change as  
$l_s$ is increased from zero:

$$\mu_e=\mu_e^{l_s=0}+\delta \mu_e.$$

 Eq.~\eqref{minimeq} can be put as
 \begin{eqnarray}
 \delta n_{B}\left(\bm{r}\right)&=& \dfrac{1}{2 \pi e \,l_{S}}\delta\phi\left(\bm{r}\right)-n_{B}^{l_{S}=0}\left(\bm{r}\right)+
  n_{e} \nonumber \\
  &  & \, \quad \forall \, \bm{r} \,\,\epsilon \,\,B
 \label{ldaref1}  
 \end{eqnarray}
where we have absorbed the Lagrange parameter in 
the constant $n_e=n_B^0+\frac{\varepsilon_{0}}{2\pi e^2 l_s}\delta
\mu_e$.  In Fourier space we get:
 \begin{eqnarray}
 \delta n_{B}\left(q\right)&=&\dfrac{1}{2 \pi e\,l_{S}R_{c}} \sum_{q^{\prime}}\delta\phi\left(q^{\prime} \right) \dfrac{ \sin
 \left[\left(q-q^{\prime}\right)
   R_{d}\right]}{\left(q-q^{\prime}\right)}\nonumber\\
&-&
 n_{B}^{l_{S}=0}\left(q\right)+n_{e}  \frac{2\sin(qR_d)}q.
 \label{eq:dnbfourier1}  
 \end{eqnarray}
Eq.~\eqref{eq:lap2d}and Eq.\eqref{eq:dnbfourier1} are a closed system 
since the quantities with $l_{S}=0$ are 
 known from the previous treatment. 

  In the case $R_{c},R_{d}>>l_{S}$ one can substitute
  $\sum_{q}\rightarrow 2R_c\int d q/(2\pi)$ and make the approximation:
  \begin{equation}
   \dfrac{2 \sin \left[\left(q-q^{\prime}\right)R_{d}\right]}{\left(q-q^{\prime}\right)}\rightarrow 2\pi \delta\left(q-q^{\prime}\right).
  \end{equation} 
Using Eq.~\eqref{eq:lap2d} we obtain:
 \begin{equation}
 \delta n_{B}\left(q\right)=\frac{2 n_{e}  \sin(qR_d)/q-
 n_{B}^{l_{S}=0}\left(q\right)}{1+1/(l_{S}|q|)} 
 \label{eq:dnbfourier}  
 \end{equation}
In the limit $q \to 0 $ we obtain $\delta n_{B}\to 0$. The uniform
component of the charge does not change and therefore
$\delta \mu_e=0$ and  $n_{e}=n_0^B$.  We have evaluated the above
expression via a discrete Fourier transform in the limit in which 
the stripes are far apart ($R_{c}/R_d\rightarrow \infty$ {\it i.e} $\nu \rightarrow 0$). This correspond
to solve the problem for a single stripe.
 The electronic density at $l_{S}=0$ can be put as:
\begin{equation*}
n_{B}^{l_{S}=0}\left(x^{\prime} \right)= 
\dfrac{2\overline n}{\pi \nu} \dfrac{1}{\sqrt{1-(x')^{2}}}
\end{equation*} 

\begin{figure}[tbp]
\includegraphics[width=8cm]{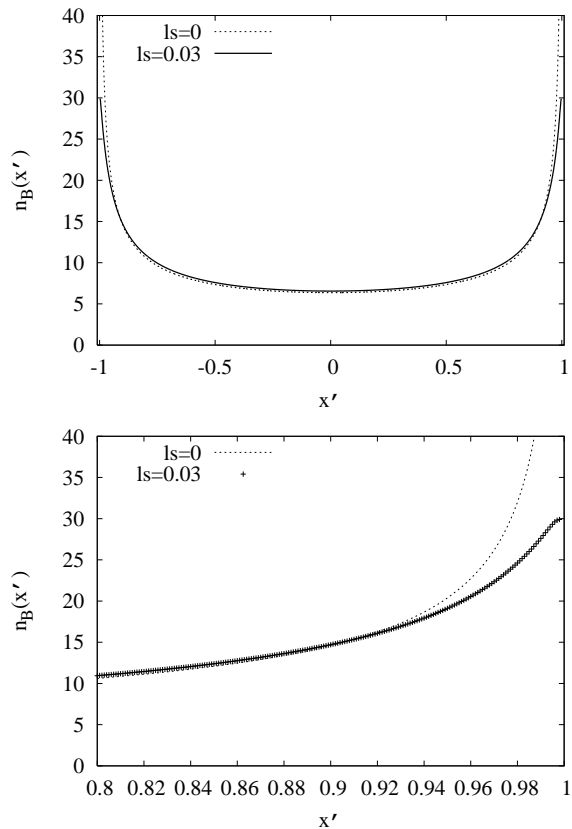}
\caption{Top panel: Comparison between the electronic charge density
  at $l_{S}/R_{d}=0$ and $l_{S}/R_{d}=0.03$ evaluated at $\nu=1$ and
$\overline n=1$. Bottom panel:Expansion near the inhomogeneity surface}
\label{ldafinale}
\end{figure}

In Fig.~\ref{ldafinale} we show the total electronic charge density
for $l_{S}/R_{d}=0.03$. 
The main difference with respect to the $l_{S}=0$ 
case is that the unphysical divergence of the density
at the stripe surface is removed
and the density tends to a finite value at the stripe boundary. Notice that at distances from
the interface larger than $l_{S}$ the charge density at $l_{S}\neq 0$ practically
coincides with the one at $l_{S}=0$ as expected.
Our previous conclusion regarding the absence of upper bounds for the  size of the domains remains
unchanged. In the entire domain the local density differs from the
average density and therefore the phase separation energy gain comes
from the whole domain and not from the electric field penetration depth as it happened in the 3D
case.

  Finally one has that the electronic potential is given by:
  \begin{equation}
  \delta
  \phi^{3D}_{e}\left(q\right)\,=-\,\dfrac{2\pi\,e}{|q|+\dfrac{1}{l_{S}}}\left[
n_{e}  \frac{2\sin(q R_d)}q
-n_{B}^{l_{S}=0}\left(q\right)\right]
  \end{equation}
  The above electrostatic potential includes the two-dimensional 
screening with $l_{S}$ as a screening parameter. It is of the same 
form of the screened potential discussed in Ref.~\onlinecite{and82}.

\section{Discussion and Conclusion}

In this work we have considered the problem of mesoscopic FPS in 2D
electronic systems frustrated by the LRC interaction and the surface
energy. In particular we concentrated on the problem of
coexistence between a 2D metal and an insulator. 

We have supposed that the inhomogeneous state is realized with two
different geometric arrangements: disks of one
phase into the other  and a state with alternating stripes of metal 
and insulator. The first arrangement comes in two different flavors:
 bubbles of insulator hosted by the metal for $\nu \sim 1$ and
metal drops hosted by the insulator for $\nu\sim 0$.

 We have defined a parameter $\lambda$ which specifies the strength 
of the frustration 
and is given by the ratio of the characteristic mixing energy to the
characteristic phase separation energy gain in the absence of
frustration. 

As in the 3D case, we have found that frustration tends to extend the global density region where the uniform metallic phase is stable. 
Within our approximation, below some critical value of $\lambda$, we have found that 
as the global density is reduced there is a second order transition from the homogeneous metallic state to an
 inhomogeneous state 
with insulating bubbles in the metal. 
Above this value of  $\lambda$ the bubble state is never 
stable and the inhomogeneous transition leads to the striped state 
with a first-order transition. Inside the inhomogeneous 
stability region by decreasing the global density first-order
``topological'' transitions induces changes in the geometry of the domains.

In Sec. \ref{secpsprop} we have analyzed the inhomogeneous state properties and we have
found that the size of inhomogeneities is not forced to have one
linear dimension smaller than the screening length $l_s$ in sharp 
contrast with the 3D case. 
 This difference stands on the qualitative 
different behavior of the screening in a 3D and in a 2D system. In the first
case the screening decays exponentially from the interfaces 
whereas in the second case it decays as a power law. 
In this last case it is possible to gain PS energy from regions far
from the skin depth.

Inhomogeneities seem to be related to a number of interesting
phenomena like colossal magnetoresistent. Our study suggests that inhomogeneities may be favored by 
engineering materials with enhanced anisotropy. For example adding insulating
layers in between metallic layers should favor mesoscopic phase
separation.

A realistic model of the 2DEG  would require considering a metallic phase with negative
compressibility as observed in the uniform phase. The
effects of disorder should also be included.\cite{ort06} Therefore 
here we limit our discussion to some generic qualitative features.  
The stabilization effect of the long-range Coulomb interaction 
has been observed  on the 2DEG where the state with
negative short-range electronic compressibility has been show to 
be stable in a certain range of densities. \cite{eis94,eis92,dul03,dul00,ila00,ila01} 
Ilani and collaborators\cite{ila00,ila01} observed that the system becomes inhomogeneous
at a mesoscopic scale when the system is at the verge of
 the metal-insulator transition suggesting a connection between the
 two phenomena. 

Another interesting finding of Ilani and
collaborators\cite{ila00,ila01} is that coming from the metallic side
the compressibility has negative spikes close to the transition. 
When integrated those spikes 
imply a decreasing step in the chemical potential as a function of
density. This behavior is similar to what we find for the transition 
from the inhomogeneous state to the metallic state as the density is
increased in the case in which the transition is first order
(c.f. Fig.~\ref{mulambda}). The main difference between our result and 
the experiment is that in our case there is a single large step at the
transition whereas Ilani {\it et al.} find many small steps around the
critical density. This behavior is easy to rationalize if one 
considers that in the presence of disorder the density will not be
uniform in all the sample. A distribution of
critical densities and fragmentation of the large step in many small
steps will be naturally produced. The minimum step size will be given by the appearance of a single drop.

We have restricted ourselves to one isolated plane but we expect the same
physics to be true for weakly coupled planes as found in some 
cuprates, nickelates manganates and other strongly correlated materials. 
As mentioned in the introduction the situation in manganites is quite
complex due to the large variety of competing phases. In layered  
cuprates one finds striped charge and spin density waves  at small
doping and a Fermi liquid at large doping. From the behavior of the
mean-field energy\cite{lor02b} it is quite natural to expect 
phase separation among the stripe state and the overdoped Fermi liquid.
To make a rough estimate of the drop size we identify the    
stripe state as the pseudogap phase and treat it as incompressible.  
If we assume that the surface energy is of order $\sigma\sim J/a$,
with $J$ the superexchange constant and $a$ the lattice constant,  
the drop radius can be written as [c.f. Eq.~\eqref{eq:rddrop}],
$$R_d\sim l_d \sim \frac{a}{\Delta x}\left(\frac {\epsilon_0 a J}{2
    a_0 Ry}\right)  $$
where $a_0$ is the Borh radius, $\Delta x\sim 0.1 $ is the range of doping 
where one expects phase separation according to MC. Using $J=0.01 Ry$,
 $a=7.2 a_0$, $\epsilon_0=5$, one gets  $2 R_d~ \sim 8 a \sim
30\AA$ which compares well with the inhomogeneities
found in Ref.\onlinecite{lan02}. This experiments reveals 
superconducting like inhomogeneities of 
roughly $50\AA$ diameter (playing the role of the compressible phase) 
embedded in insulating like regions. A similar rough 
estimate gives $\lambda\sim 1$.

The constraint on the maximum size of inhomogeneities 
in 3D systems makes the conditions for stability of a 
phase separated state very stringent.\cite{lor01I} In 2D system instead 
we have found that this constraint does not apply and frustrated phase separation at the mesoscopic scale 
is much more favorable.  
 This may be the reason why inhomogeneities are often found in
quasi two-dimensional electronic systems.

\acknowledgments 
C. Ortix acknowledges partial support from 
INFM and 
Dipartimento di Fisica, Universit\`a di Roma ``La Sapienza'' during
part of this work and M. Beccaria for useful discussions.

\appendix
\section{Electrostatic energy in the uniform density approximation}
\label{secappe}
Here we compute the explicit expressions of the electrostatic energy
within the UDA for the two geometries considered.  We start by
considering regular arrays of inhomogeneities and dividing
the system in Wigner-Seitz cells in such a way that each cell is
globally neutral. The electrostatic energy is $ N_{d}$ 
times the Coulomb energy of one cell.  

As in Ref.~\onlinecite{lor01I} we take the density profile 
inside each cell as
$-e\left(n_{B}\,-\,n_{A}\right)$ inside the B-phase inhomogeneity 
of radius $ R_{d} $ compensated by the background charge density
$e\left(n-n_{A}\right)$.  

In 3D the computation of the Coulomb energy is facilitated
by the use of Gauss theorem.\cite{lor01I} In the present 2D case 
with the full 3D Coulomb interaction 
Gauss theorem is not useful so we use the explicit expression:\cite{jac75}
\begin{equation}
e_{el}\,=\,\frac{1}{2\,\varepsilon_{0}}\,\dfrac{1}{V_{d}}\int_{V_{d}}\, d^{2}\bm{r}\,
\left(\rho_{d}+\rho_{b}\right)\,\left(\phi_{d}\left(\bm{r}\right)+
\phi_{b}\left(\bm{r}\right)\right)
\label{coulombenergy}
\end{equation}
where $ V_{d} $ is the 2D volume of one cell, 
$ \rho_{d}\equiv -e\left(n_{B}\,-\,n_{A}\right) $
is the inhomogeneity  charge density 
and $\rho_{b}=e\left(n-n_{A}\right) $ is the effective background 
charge density. 
We need to evaluate the Coulomb  potential $\phi_{d}$ 
generated by the inhomogeneities of all cells  and 
the Coulomb potential $\phi_{b}$  generated by
the whole background charge density. The latter corresponds to the 
Coulomb potential of an infinite uniformly charged plane and can be
taken as constant in the plane. Because of the global neutrality the value of
the constant does not affect the electrostatic energy and we can take
 $\phi_{b}=0$. 

The inhomogeneity Coulomb potential
is the sum of the $ N_{d} $ contributions from each cell.  
We separate the contribution of the cell where we are integrating from the contribution of the other cells. In this way 
Eq.~(\ref{coulombenergy}) separates in a self-energy contribution 
($e_{el}^{\Sigma}$) and an interaction contribution $e_{el}^{int}$:
\begin{eqnarray}
e_{el}^{\Sigma}\,&=&\,\frac{V_{d}^{-1}}{2\,\varepsilon_{0}}\,\int_{V_{d}}\, d^{2}\bm{r}\,
\left(\rho_{d}+\rho_{b}\right)\phi^{\Sigma}\left(\bm{r}\right)
\label{selfcoulomb} \\
e_{el}^{int}\,&=&\,\frac{V_{d}^{-1}}{2\,\varepsilon_{0}}\,\int_{V_{d}}\, d^{2}\bm{r}\,
\left(\rho_{d}+\rho_{b}\right)\,\phi^{int}\left(\bm{r}\right)
\label{intercoulomb}
\end{eqnarray}

\subsection{Stripe geometry}
In this geometry we assume that the system is divided in cells of
width $ 2\,R_{c} $ and length $ L $ with periodic boundary
conditions. Within each
cell the width of the inhomogeneity with charge density $ \rho_d$
 is equal to $ 2\,R_{d} $. 

To compute the electrostatic energy we
have evaluated the two expressions 
Eqs.~\eqref{selfcoulomb},~\eqref{intercoulomb}.
The interaction potential $\phi^{int}$ has been numerically computed
truncating the sum of the contribution from each cell to a finite
number of cells: $N_{c}$. For $N_c$ not too large the Coulomb energy
is asymmetric. However for $N_c$ of the order of $10^{2}$ the Coulomb
potential becomes symmetric respect to the "phase-exchange" symmetry 
($A\rightarrow B$ , $\nu\rightarrow 1-\nu$) indicating the achieved 
convergence.

In the limit $ \nu\rightarrow 0 $, 
$R_{d}<<R_{c}$,  $\phi^{int}$ can be neglected
 so that the total electrostatic energy is well approximated in this
limit by:
\begin{equation}
\lim_{\nu\rightarrow 0} e_{el}^{\Sigma}= \frac{e^{2}}{\varepsilon_{0}}\left(n_{B}-n_{A}\right)^{2}R_{c}\,2\,\nu^{2}\left[-\ln
\nu\right]
\end{equation}
We can obtain a similar approximation in the opposite 
limit $\nu\rightarrow 1$ by imposing the 
''phase-exchange'' symmetry. This leads to the approximate
expression Eq.~\eqref{electrostripes} which interpolates between the 
$\nu\rightarrow 0,1 $ limits. 
The comparison between the
approximate and the numerical result is shown in
Fig.~\ref{ucomparison}. 
in which are reported the $u$ functions considering the different
electrostatic energies. We see that the approximation to keep only the
self energy term is indeed very good. 

\subsection{Drop geometry}
From the computation in the stripe geometry it is clear that in the
limit of $\nu\rightarrow 0$  and $\nu\rightarrow 1$
the dominant electrostatic term is the self-energy.  Since the drops
are zero dimensional objects we expect that the effect of the 
intercell  terms to be even smaller than for the case of
 stripes. To check this hypothesis we first estimate the interaction energy between
 the cells in this way: if we look to the total system of $ N_{d} $
cells in the limit of small volume fraction,
 the drops can be considered as negative point-charges at distance  
$R_{c} $ since $\nu\rightarrow 0$ is equivalent to
$R_{d}<<R_{c}$ and we assume that the drops arrange in a Wigner
crystal. The electrostatic energy is given by:\cite{wig34}
\begin{equation}
E_{el}^{int}\,\propto \,-\alpha
\,\frac{\left[e\,R_{d}^{2}\,\left(n_{B}-n_{A}\right)\right]^{2}}{R_{c}}
\end{equation}
Since $\nu=R_{d}^{2}/R_{c}^{2}$ for the drop geometry and referring to
the energy per unit volume one has:
\begin{equation*}
e_{el}^{int}\propto\,
\frac{e^{2}}{\varepsilon_{0}}\left(n_{B}-n_{A}\right)^{2}R_{c}\,\alpha
\nu^{2}
\end{equation*}

The Coulomb self-energy of a cell of radius $ R_{d} $ can
be evaluated noting that the cell charge density in units of $(-e) $
can be written in the Fourier space as \cite{shi99}:
\begin{equation*}
\delta n\left(k\right)=2\pi\left(n_{B}-n_{A}\right)R_{d}^2
\left(\frac{J_{1}\left(kR_{d}\right)}{k R_{d}}-\frac{J_{1}\left(k
R_{c}\right)}{k R_{c}}\right)
\end{equation*}
where $\delta
n\left(k\right)=\left[n_{B}\left(k\right)-n_{A}\left(k\right)\right]$
.  The electrostatic energy per unit volume is:
\begin{equation*}
e_{el}^{\Sigma}\,=\,\frac{e^{2}}{2\,\varepsilon_{0}\,\pi\,R_{c}^{2}}\,\int\,\frac{d^{2}k}{\left(2\,\pi\right)^{2}}\,
\left(n_{B}\left(k\right)-n_{A}\left(k\right)\right)^{2}\,\frac{2\,\pi}{k}
\end{equation*}
Computing the coulombic self-energy per unit volume in the
Fourier-space one obtains:
\begin{equation}
e_{el}^{\Sigma}=\frac{e^{2}}{\varepsilon_{0}}\left(n_{B}-n_{A}\right)^{2}R_{c}\,
\frac{8}{3}\left[\nu^{\frac{3}{2}}\,+\,o\left(\nu^{2}\right)\right]
\end{equation}
were we have kept the dominant contribution when  $\nu\rightarrow 0$.
Also for drops one obtains that in this limit 
the self-energy term dominates the electrostatic energy.   

The limit $\nu\,\rightarrow\,1$ can be obtained by replacing  
$\nu\rightarrow 1-\nu$.  This leads to the approximate
electrostatic energy expression Eq.~\eqref{electrodrops}
which as for stripes interpolates between the two limits. 

Notice that since the inter drop interaction is negligible our
computation is independent of the lattice structure of the crystal and
is also valid for an amorphous configurations of drops.

\section{Limit of small screening length in the UDA and comparison between the UDA and the LDA}
\label{seckbinf} 
To compare the LDA and the UDA we use the particularly simple limit $l_{S}\rightarrow 0$. 
 Formally this can be achieved by making $\kappa_{B}\rightarrow
\infty$ so that the 
$B$ phase bulk energy becomes a linear function of density.
If $n_B^0$ is kept constant the line intersects the energy of the A-phase in this limit. 

We will instead consider the energy density difference $\Delta E_{g}$ 
between the $B$ and the $A$ phase at zero density fixed. In this case the MC density 
$n_{B}^{0}\,=\,[\Delta E_{g} \kappa_{B}]^{1/2}$ diverges.

To make things less abstract we can consider the following example:
classical electrons with a short range attractive interaction in the
lattice and at low temperature. The attraction stabilizes a crystal 
at high density. The energy per unit volume of the uniform
crystal is $f_A^0=-z v n_0/2$ 
where $z$ is the coordination number $v$ is the short range attraction
and $n_0$ is the density of the incompressible crystal phase. 

For small global density the attraction can be neglected. 
Electrons form a uniform ``metallic''
phase. Since the electrons are classical (no tight binding hopping term)
  the  chemical potential becomes independent of the density 
(thus the compressibility is infinite) and taken to be zero.

In the absence of long-range
 interaction and for electronic densities $n_e<n_0$ 
this system phase separate into the high density crystal phase and the
zero density empty phase.  In the presence of the 
long-range Coulomb interaction separation between the high-density
crystal phase and a low-density metallic phase becomes possible.

In order to keep notations consistent with the previous section 
we consider the hole density $n=n_0-n_e$. The incompressible A
phase is then at $n=0$ and has energy  $f_A^0<0$
  and the infinitely compressible metallic phase has $n>0$ and energy 
 $f_B(n)=0$.

The free energies of the homogeneous state and the  
PS-states (drop and stripes geometries)  can be put as:
 \begin{eqnarray} 
 f_{H}\left(n\right)&=&0 \\
 f_{D,S}\left(n,\nu \right)&=&(1- \nu) f_A^0+\left(\frac{\sigma
   e^{2}}{\varepsilon_{0}}\right)^{\frac{1}{2}} n \frac{
   u\left(\nu\right)}{\nu} \label{eq:fdsuda} 
 \end{eqnarray}

Minimizing this expression respect to $\nu$ one obtains the 
optimum $\nu$ value. 

Now in order to test the accuracy of the UDA 
we will compare this approximation with the LDA derived in Sec. \ref{seclda}. 
For simplicity we
restrict to the case of separation between and 
incompressible phase  and an infinitely compressible phase.

The energy can be evaluated in the 
LDA using the  spatial dependence of the electron density 
found in Sec.~\ref{seclda} for $l_s=0$ 
[c.f. Eq.~\eqref{densitalocale0ordine}]. 

As discussed above we assume that $f_A$ is a negative constant
and $$f_{B}\left[n_{B}\left(\bm{r}\right)\right]\,= 0  $$
With these conventions the LDA free energy  functional reads:
\begin{equation}
F\,=\,V_{A}\,f_{A}^{0}+ \sigma\,\Sigma_{A B} + \dfrac{e}{2\,\varepsilon_{0}}\,\int_{V_{A}} \overline n\,\phi^{l_{S}=0}\left(\bm{r}\right)\,d^{2}\bm{r}
\end{equation}
where we have used that in the metallic phase domains the electrostatic 
potential for $l_{S}=0$ is constant and thus give no contribution to the 
Coulomb energy.

For the stripes geometry the energy density reads:
$$
f_S=\left(1-\nu\right)\,f_{A}^{0}+ \frac{e^{2}}{4\varepsilon_{0}}\,\overline n^{\,2}\, R_{c}\,
\frac{u_{LDA}\left(\nu\right)^2}{\nu^{2}}+\dfrac{\sigma}{R_{c}}.
$$
Here $u_{LDA}$ is given in terms of the potential 
at $l_{S}=0$:
$$
u_{LDA}\left(\nu\right)^2=
\dfrac{ 2\nu^2}{e\,\overline n R_{c}} \int_{\nu}^{1}
\phi^{l_{S}=0}\left(\tilde{x}\right) d\tilde{x}
$$
where $\tilde{x}$ is the dimensionless coordinate $x/R_{c}$ and
the potential at $l_{S}=0$ reads:

\begin{equation}
\phi^{l_{S}=0}\left(\tilde{x}\right)\,=\,4 R_{c}\, e\, \overline{n}
\cosh^{-1}{\dfrac{\sin{|\dfrac{\pi \tilde{x}}{2}}|}{\sin{\dfrac{\pi
        \nu}{2}}}} 
\end{equation} 

Minimizing the mixing energy $e_{m}=e_{el}+e_{\sigma}$ respect to the
cell radius $R_{c}$ one finally finds:
\begin{equation*}
f_{S}\left(\overline n , \nu\right)=\left(1-\nu\right) f_{A}^{0}+ \left(\frac{\sigma
  e^{2}}{\varepsilon_{0}}\right)^{\frac{1}{2}}\, \overline n
\frac{u_{LDA}\left(\nu\right)}{\nu} 
\end{equation*} 
The difference of energy between the PS and 
 the homogeneous states can be put in the same form as 
in the UDA Eq.~\eqref{eq:fdsuda}.  
The only difference with the UDA is encoded in the 
 function $u_{LDA}$. In Fig.~(\ref{ucomparison}) we compare
the two geometrical $u$ function parameterizing the mixing energies. The LDA 
function implies a lower mixing energy since we are relaxing the
uniform density constraint. This relaxation energy gain however is
 small demonstrating the accuracy of the UDA approximation for
 thermodynamic quantities as it was found also in the 3D case.\cite{lor01I}


\begin{thebibliography}{10}

\bibitem{mul92}
 in {\em Phase separation in cuprate superconductors}, edited by K.~A. Muller
  and G. Benedek (World Scientific, Singapore, 1992).

\bibitem{sig93}
 in {\em Phase separation in cuprate superconductors}, edited by E. Sigmund and
  K.~A. Muller (Springer-Verlag, Berlin, 1993).

\bibitem{mor99}
A. Moreo, S. Yunoki, and E. Dagotto, Science {\bf 283},  2034  (1999).

\bibitem{nag83}
E.~L. Nagaev, {\em {Physics of magnetic semiconductors }} (MIR, Moscow, 1983).

\bibitem{nag98}
E.~L. Nagaev, A.~I. Podel'shchikov, and V.~E. Zil'bewarg, J. Phys.: Condens.
  Matter {\bf 10},  9823  (1998).

\bibitem{hen98}
M. Hennion, F. Moussa, G. Biotteau, J. Rodriguez-Carvajal, L. Pinsard, and A.
  Revcolevschi, Phys.\ Rev.\ Lett. {\bf 81},  1957  (1998).

\bibitem{low94}
U. L\"ow, V.~J. Emery, K. Fabricius, and S.~A. Kivelson, Phys.\ Rev.\ Lett.
  {\bf 72},  1918  (1994).

\bibitem{cas95b}
C. Castellani, C. {Di Castro}, and M. Grilli, Phys.\ Rev.\ Lett. {\bf 75},
  4650  (1995).

\bibitem{ila00}
S. Ilani, A. Yacoby, D. Mahalu, and H. Shtrikman, Phys.\ Rev.\ Lett. {\bf 84},
  3133  (2000).

\bibitem{ila01}
S. Ilani, A. Yacoby, D. Mahalu, and H. Shtrikman, Science {\bf 292},  1354
  (2001).

\bibitem{pan01}
S.~H. Pan, J.~P. O'neal, R.~L. Badzey, C. Chamon, H. Ding, J.~R. Engelbrecht,
  Z. Wang, H. Eisaki, S. Uchida, A.~K. Gupta, K.-W. Ng, E.~W. Hudson, K.~M.
  Lang, and J.~C. Davis, Nature (London) {\bf 413},  282  (2001).

\bibitem{mce03}
K. {McElroy}, R.~W. {Simmonds}, J.~E. {Hoffman}, D.-H. {Lee}, J. {Orenstein},
  H. {Eisaki}, S. {Uchida}, and J.~C. Davis, nat {\bf 422},  592  (2003).

\bibitem{lan02}
K.~M. Lang, V. Madhavan, J.~E. Hoffman, E.~W. Hudson, H. Eisaki, S. Uchida, and
  J.~C. Davis, nat {\bf 415},  412  (2002).

\bibitem{bec02}
T. Becker, C. Streng, Y. Luo, V. Moshnyaga, B. Damaschke, N. Shannon, and K.
  Samwer, Phys.\ Rev.\ Lett. {\bf 89},  237203  (2002).

\bibitem{sal01}
M.~B. Salamon and M. Jaime, Rev.\ Mod.\ Phys. {\bf 73},  583  (2001).

\bibitem{per97}
N.~M.~R. Peres, J.~M.~P. Carmelo, D.~K. Campbell, and A.~W. Sandvik, Z.\ Phys.\
  B {\bf 103},  217  (1997).

\bibitem{lar01}
S. Larochelle, A. Mehta, N. Kaneko, P.~K. Mang, A.~F. Panchula, L. Zhou, J.
  Arthur, and M. Greven, Phys.\ Rev.\ Lett. {\bf 87},  095502  (2001).

\bibitem{zha02}
L. Zhang, C. Israel, A. Biswas, R.~L. Greene, and A. de~Lozanne, Science {\bf
  298},  805  (2002).

\bibitem{lar05}
S. Larochelle, A. Mehta, L. Lu, P.~K. Mang, O.~P. Vajk, N. Kaneko, J.~W. Lynn,
  L. Zhou, and M. Greven, Phys.\ Rev.\ B {\bf 71},  024435  (2005).

\bibitem{fat99}
M. F{\"a}th, S. Freisem, A.~A. Menovsky, Y. Tomioka, J. Aarts, and J.~A.
  Mydosh, Science {\bf 285},  1540  (1999).

\bibitem{ren02}
C. Renner, G. Aeppli, B.-G. Kim, Y.-A. Soh, and S.-W. Cheong, Nature (London)
  {\bf 416},  518  (2002).

\bibitem{kra94}
S.V.Kravchenko, G.V.Kravchenko, J.E.Furneaux, V.M.Pudalov, and M. D'Iorio, Phys.\
  Rev.\ B {\bf 50},  8039  (1994).

\bibitem{kra95}
S.V.Kravchenko, W.E.Mason, G.E.Bowker, J.E.Furneaux, V.M.Pudalov, and M. D'Iorio,
  Phys.\ Rev.\ B {\bf 51},  7038  (1995).

\bibitem{sim98}
M.Y.Simmons, A.R.Hamilton, M.Pepper, E.H.Linfield, P.D.Rose, D.A.Ritchie, A.K.Savchenko, and T.G. Griffiths,
  Phys.\ Rev.\ Lett. {\bf 80},  1292  (1998).

\bibitem{eis94}
J.P.Eisenstein, L.N.Pfeiffer, and K.W.West, Phys.\ Rev.\ B {\bf 50},  1760
  (1994).

\bibitem{eis92}
J.P.Eisenstein, L.N.Pfeiffer, and K.W.West, Phys.\ Rev.\ Lett. {\bf 68},  674
  (1992).

\bibitem{dul03}
S.C.Dultz and H.W.Jiang, J. Phys. Soc. Jpn. {\bf 72},  674  (2003).

\bibitem{dul00}
S.C.Dultz and H.W.Jiang, Phys.\ Rev.\ Lett. {\bf 84},  4689  (2000).

\bibitem{seu95}
M. Seul and D. Andelman, Science {\bf 267},  476  (1995).

\bibitem{lor93prl}
C.~P. {Lorenz}, D.~G. {Ravenhall}, and C. J.{Pethick}, Phys.\ Rev.\ Lett. {\bf
  70},  379  (1993).

\bibitem{fra99}
P. Fratzl, O. Penrose, and J.~L. Lebowitz, J. of Stat. Phys. {\bf 95},  1429
  (1999).

\bibitem{kit46}
C. Kittel, Phys.\ Rev. {\bf 70},  965  (1946).

\bibitem{lan84}
L. Landau and E. Lifshitz, {\em Electrodynamics of Continuous Media} (Pergamon,
  New York, 1984).

\bibitem{lor01I}
J. Lorenzana, C. Castellani, and C. {Di Castro}, Phys.\ Rev.\ B {\bf 64},
  235127  (2001).

\bibitem{lor01II}
J. Lorenzana, C. Castellani, and C. {Di Castro}, Phys.\ Rev.\ B {\bf 64},
  235128  (2001).

\bibitem{lor02}
J. Lorenzana, C. Castellani, and C. {Di Castro}, Europhys. Lett. {\bf 57},  704
   (2002).

\bibitem{mur02}
C.~B. Muratov, Phys.\ Rev. \ E {\bf 66},  066108  (2002).

\bibitem{note}
Throughout this work we call a metal a charged fluid which is compressible
  regardless of transport properties except when we obviously refer to
  transport as in "metal-insulator transition".

\bibitem{spi03}
B. Spivak, Phys.\ Rev.\ B {\bf 67},  125205  (2003).

\bibitem{spi04}
B. Spivak and S.~A. Kivelson, Phys.\ Rev.\ B {\bf 70},  155114  (2004).

\bibitem{ort06}
C. Ortix, J. Lorenzana, and C. {Di Castro}, in preparation (unpublished).

\bibitem{eme90c}
V.~J. Emery, S.~A. Kivelson, and H.~Q. Lin, Phys.\ Rev.\ Lett. {\bf 64},  475
  (1990).

\bibitem{kiv90}
S.~A. Kivelson, V.~J. Emery, and H.~Q. Lin, Phys.\ Rev.\ B {\bf 42},  6523
  (1990).

\bibitem{can91}
N. Cancrini, S. Caprara, C. Castellani, C. {Di Castro}, M. Grilli, and R.
  Raimondi, Europhys. Lett. {\bf 14},  597  (1991).

\bibitem{gri91}
M. Grilli, R. Raimondi, C. Castellani, C. {Di Castro}, and G. Kotliar, Phys.\
  Rev.\ Lett. {\bf 67},  256  (1991).

\bibitem{don95}
P.~G.~J. van Dongen, Phys.\ Rev.\ Lett. {\bf 74},  182  (1995).

\bibitem{oka00}
S. Okamoto, S. Ishihara, and S. Maekawa, Phys.\ Rev.\ B {\bf 61},  451  (2000).

\bibitem{lor02pb}
J. Lorenzana, C. Castellani, and C. {Di Castro}, Physica B {\bf 320},  56
  (2002).

\bibitem{and82}
T. Ando, A.B.Fowler, and F.Stern, Rev.\ Mod.\ Phys. {\bf 54},  437  (1982).

\bibitem{tin75}
M. Tinkham, {\em Introduction to superconductivity} (McGraw-Hill, New York,
  1975).

\bibitem{smy68}
W.R.Smythe, {\em Static and dynamic electricity} (McGraw-Hill, New York, 1975).

\bibitem{lor02b}
J. Lorenzana and G. Seibold, Phys.\ Rev.\ Lett. {\bf 89},  136401  (2002).

\bibitem{jac75}
J.D.Jackson, {\em Classical electrodynamics} (J.Wiley and sons, New York,
  1975).

\bibitem{wig34}
E. Wigner, Phys.\ Rev. {\bf 46},  1002  (1934).

\bibitem{shi99}
J. Shi, S. He, and X.C. Xie, Phys.\ Rev.\ B {\bf 60},  R13950  (1999).

\end{thebibliography}
\end{document}